\documentclass{article}
\usepackage{spconf,amsmath,graphicx}
\usepackage[table,xcdraw]{xcolor}
\usepackage{setspace}
\makeatletter
\def\@name{Jasper Kirton-Wingate$^{\dagger}$, Shafique Ahmed$^{\star}$, Adeel Hussain$^{\dagger}$, Mandar Gogate$^{\dagger}$, Kia Dashtipour$^{\dagger}$,\\ Jen-Cheng Hou$^{\star}$, Tassadaq Hussain$^{\dagger}$, Yu Tsao$^{\star}$, Amir Hussain$^{\dagger}$}
\makeatother
\address{$^{\dagger}$Edinburgh Napier University, $^{\star}$Academia Sinica}
\title{Towards Environmental Preference Based Speech Enhancement For Individualised Multi-Modal Hearing Aids}

\begin{document}
\maketitle
\onecolumn

\begin{abstract}
Since the advent of Deep Learning (DL), Speech Enhancement (SE) models have performed well under a variety of noise conditions. However, such systems may still introduce sonic artefacts, sound unnatural, and restrict the ability for a user to hear ambient sound which may be of importance. Hearing Aid (HA) users may wish to customise their SE systems to suit their personal preferences and day-to-day lifestyle. In this paper, we introduce a preference learning based SE (PLSE) model for future multi-modal HAs that can contextually exploit audio information to improve listening comfort, based upon the preferences of the user. The proposed system estimates the Signal-to-noise ratio (SNR) as a basic objective speech quality measure which quantifies the relative amount of background noise present in speech, and directly correlates to the intelligibility of the signal. Additionally, to provide contextual information we predict the acoustic scene in which the user is situated. These tasks are achieved via a multi- task DL model, which surpasses the performance of inferring the acoustic scene or SNR separately, by jointly leveraging a shared encoded feature space. These environmental inferences are exploited in a preference elicitation framework, which linearly learns a set of predictive functions to determine the target SNR of an AV (Audio-Visual) SE system. By greatly reducing noise in challenging listening conditions, and by novelly scaling the output of the SE model, we are able to provide HA users with contextually individualised SE. Preliminary results suggest an improvement over the non-individualised baseline model in some participants. %Additionally, other related effects of the noise suppression are discussed. %As a result, the proposed dataset allows the user to engage in a VR experience of two-target speaker listening, whilst enabling the evaluation of multi-modal SE models within a VR HA context. The sentences chosen to be spoken were taken from British IEEE Harvard sentences. The sentences are employed in a turn-taking scenario and a more challenging sentence overlap scenario with simultaneous speech. The target speaker can be chosen by the head orientation angle, or by other measures that predict attention, such as eye tracking. Other sensors can also be employed to assess the LE effects of aiding through VR.
        %My research aims to investigate subjective barriers to uptake of modern SE algorithms, and further to this, help individuals take control of Audio Artificial Intelligence, to make it work best for them and the situations they find themselves in.   

\end{abstract}


\section{Introduction}\label{sec:intro}
Speech enhancement (SE) models are typically evaluated by criteria that objectively measure both the speech quality and intelligibility (i.e. PESQ, STOI). Ideally, when evaluating for specific application in HA's, the performance of SE should be benchmarked after modelling HA signal degradation and enhancement (i.e. HASQI, HASPI) \cite{haspi-hasqi}, as well as computational complexity and it's subsequent effect on latency of the end-to-end output signal transformation. Additionally, whilst SE models may perform well on simulated, synthetic data, it is of critical importance to evaluate them in multi-channel, realistic, audio-visual scenarios to test ecological efficacy.

However, the ideal SE model is not necessarily `one size fits all'. In established research, it has been posited that preference for the amount of Noise Reduction (NR) for various signal-to-noise ratios (SNRs) differs amongst hearing aid (HA) users~\cite{Neher2016}.
Moreover, given the scientific description of the highly individualised and non-linear pathology that constitutes hearing loss~\cite{LESICA2018174,diag-noise-induced-hearing-loss,fereczkowski2024amplitude}, there have been relatively few attempts to personalise modern, non-linear SE algorithms with respect to the hearing impaired listeners preferences. This gap has been highlighted recently in~\cite{creating-clarity-dl}, describing the inferred best direction of future hearing aid research.

\begin{figure}
\centering
\includegraphics[scale=0.5]{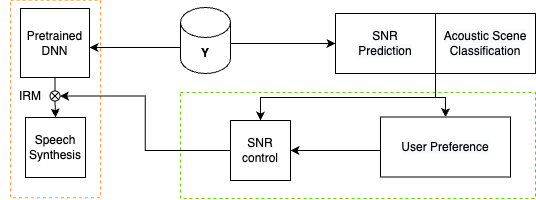} \caption{PLSE HA System Diagram with Experimental Dataset Y (GRID-CHIME3)}\label{fig:PLSE} %the clear arrows represent test-time-only processing, note that the ASPIRE dataset is omitted for first round of experiments}
    
\end{figure}

In this paper, we propose a framework for individualised AVSE, as shown in Fig.~\ref{fig:PLSE}, that controls the output of DL-based AV SE models according to the user preference of NR at different levels of background noise within various acoustic environments. In the subjective evaluation section, Normal Hearing (NH) and Hearing Impaired (HI) participants are tested and their preferences for NR, alongside the subsequent evaluation of the PLSE and baseline model, are analysed. We evaluate the resultant noise and speech quality alongside the speech intelligibility from the PLSE in comparison to the noisy condition and the baseline model.
 %(and in some cases improving it). 
Simultaneously, by building on previous work \cite{jkw-amhat2023}, a multi-task DL-based SNR estimation and Acoustic Scene Classification (ASC) model is designed to hierarchically model and predict the individualised preference for AVSE, which builds a foundational system to provide PLSE in real test conditions. Once the user's preference function has been estimated after preference elicitation, we hypothesise that it can be utilised to achieve better overall listening than using a static, `one-size-fits-all' NR model in HA, without significantly impeding the intelligibility.

The rest of this paper as organised as follows: Section 2 reviews related work, Section 3 introduces our proposed PLSE framework including the multi-task ASC \& SNR prediction architecture, Section 4 introduces the experimental set up for the subjective results from NH and HI participants presented in Section 5, and finally Section 6 presents a general discussion with references to related literature, limitations of this study, concluding remarks and future work directions.

\section{Related Work}

The following section describes related research in the following parts: Hearing Aid Personalisation, Speech Enhancement, Performance Assessment \& SNR Prediction, Acoustic Scene Classification and Multi-Task Learning.

\subsection{Hearing Aid Personalisation}

Within the HA personalisation literature there are many studies and models for fine-tuning, particularly in individualising the parameters in digital multi-band dynamic range compression algorithms that exist in most modern HAs, initially set according to prescriptions such as the clinical standard NAL-NL2 \cite{keidser2011nal,casolani2024evaluation}. By further adjusting the patient's audiogram based prescription and/or compression parameters based on personal preference, we see a significant increase in user satisfaction of the audio quality \cite{nielsen2013hearing, gp-personal}. More recently, Drakopoulos et al.~\cite{drakopoulos_differentiable_2022} %(replace with journal citation) 
proposed to attempt to invert an individual auditory pathology by decreasing the error between Normal Hearing (NH) and Hearing Impaired (HI) simulated auditory nerve responses using Deep Learning. %\cite{HA-SSL} used a Self-Supervised framework.

\subsubsection{Individualised Speech Enhancement}

In terms of individualised SE, Bhat et al.~\cite{Bhat2018} proposed a formant based SE framework to customise the noise suppression and subsequent speech distortion according to the user preference elicited via a smartphone based elicitation system. The model modulates the frequency domain formants to control the HA output based on user preferences whilst attempting to maintain speech intelligibility. Other studies have focused on personalising SE models with respect to the listener's preference by fine-tuning NR parameters~\cite{Bhat2018}, or with respect to the audiogram via spectral change enhancement~\cite{Chen2018}. However, these approaches do not utilise recent advances in Deep Learning technologies, which have significantly improved intelligibity and quality over other approaches~\cite{creating-clarity-dl} \cite{sixty-years-se} (most frequently demonstrated by objective metrics such as PESQ and STOI). %However, 
%There is also evidence that for particular algorithms, NR preference varies with noise SP (cite). 
% this sentence needs more work
% Notable work has focused on personalising formant based SE with respect to user preferences for the approximate SNR \cite{Bhat2018}, however, 
%to the best of our knowledge there are no attempts in the literature to solve the aforementioned issue for audio or audio-visual DL based SE (AVSE).
Moreover, the differential in preferences that are shown for varying types of noise experienced in the real world (that are here treated as acoustic scenes), whilst accomodating for the trade off between preference for noise level and naturalness~\cite{percep-fx-nr,kubiak_relation_2022}, has not been explicitly accounted for in adaptive DL based SE. Furthermore, rich feature embeddings that are gained from DL ASC models, have not been investigated in confluence with this differential in preference, according to a hierarchy of the Acoustic Scene and SNR.    

\subsection{Speech Enhancement}
\subsubsection{Speech Enhancement Performance Assessment \& Intelligibility Prediction}

Aside from simply measuring the SNR, it is possible to predict the intelligibility of the signal: either as a general population function (e.g STOI), or as a function of the individual's hearing loss and hearing aid (e.g. measured by Audiogram) (HASPI, HASQI \cite{haspi-hasqi}). Though these may offer some advantage over predicting the SNR as they are more related to speech perception, SNR is a higher level feature and thus provides more generalisation for analyses. %However in this study, we predict the noise type alongside the frequency weighted Segmented SNR which places higher importance on intelligible frequencies of the signal, and thus correlates more to the intelligibility (cite). 
It is thought that this will add rich enough information to correlate with the preference, whilst providing data that can be compared with in-clinic SIN tests. %(cite). %Practically speaking, SIN test results could act as a prior to inform the characteristics of the PLSE system, which is something we are exploring in future work.

To achieve a reasonable estimate of the SNR computationally, generally the Segmented SNR is calculated, which takes the SNR at particular spectral time windows utilising the STFT and then averages the values over an extended length of time \cite{segsnr}. %To increase the correlation with the intelligibility of the signal, it is possible to use the Frequency Weighted Segmented SNR (cite).
In addition to this methods also exist for predicting the intelligibility of the signal utilising Deep Learning. %last bit necessary? 
Several DL-based assessment tools have been developed utilising a range of model architectures, e.g., BiLSTM \cite{Quality-net}, CNN \cite{feng_nonintrusive_2022}, and CNN-BiLSTM \cite{mosnet}. Additionally, attention mechanisms \cite{stoinet} and multitask learning \cite{multi-obj-speech-assess} have also been employed to enhance assessment abilities. 

\subsubsection{Audio-Visual Speech Enhancement}

Afouras \cite{afouras2018conversation} proposed a deep AVSE network capable of isolating a speaker's voice based on the lip region in the corresponding video input. This is achieved by predicting both the magnitude and phase of the target signal. The Lip Reading in the Wild 2 (LRS2) dataset has been utilized to evaluate the performance of this approach. The experimental results demonstrate that the model excels at isolating real-world challenging examples. Additionally, Wang et al. \cite{wang2020robust} presented an online approach that predicts rough speech patterns from the visual stream. These predictions are then integrated into the output of an AV-based system, resulting in a conservative yet robust utilisation of the weak information embedded in the visual stream. Initial experimental results show that the proposed method outperforms the audio-only baseline at different Signal-to-Noise Ratio (SNR) levels. More recently, Chern et al. \cite{chern2023audio} introduced a method leveraging the pre-trained AV-HuBERT model followed by an SE module for AVSE and Audio-Visual Speaker Separation (AVSS). Comparative experimental results confirm the effectiveness of the model with a fine-tuning strategy, demonstrating that the multimodal self-supervised embeddings obtained from AV-HuBERT can be generalized to AV regression tasks.

The SE, or more specifically, NR algorithm employed in this study is a class of DL based AV SE \cite{michelsanti_overview_2021}. Intelligibility-Oriented AV SE (IOAVSE)~\cite{io-avse} in-particular has been chosen because of it's high performance in comparison with other NR algorithms at low SNR, as well as it's Ideal Ratio Mask (IRM) based output which makes adjustment of the NR `target SNR' ($SNR^*$) relatively simple, via the modulation of the activation function at the final output layer.

\subsection{Acoustic Scene Classification}

Acoustic Scene Classification (ASC) is a sub-field of machine listening that deals with the classification and representation learning of acoustic scenes or environments. %Early approaches included (CITE CITE). 
Recently the DCASE challenges have provided an opportunity for international researchers to improve the SOTA in ASC. Deep Learning based models, utilising convolutional, recurrent architectures have been utilised to achieve SOTA results \cite{dl_asc_review}. Furthermore, it has recently been of interest to generate interpretable semantic sound recognition models that enable meaningful navigation of the encoded feature space \cite{semantically-informed-deep-neural-networks}. In this paper, we additionally make contribution by extending this idea to acoustic scene recognition to make it interpretable both for modelling and in potential future work, clinical use via preference learning.

Within the context of HA's, there are a number of factors to consider when deploying an ASC system, such as Real-time factor, microphone placement, directionality and interference, and power saving efficiency. \cite{martin2022low} presents the results of the DCASE 2022 challenge for low complexity ASC with devices. Additionally \cite{panariello2022low} explores this within the context of hearing aids.

\subsection{Multi-Task Learning}
Multi-task learning (MTL) has emerged as a prominent paradigm in machine learning, aiming to improve the performance of multiple related tasks simultaneously by leveraging their shared information or dependencies. Rather than training individual models for each task independently, MTL frameworks exploit the inherent relationships among tasks to enhance generalization and learning efficiency.

The resurgence of neural networks has fueled the exploration of multitask learning in deep learning frameworks. \cite{MTL} provided a comprehensive survey of multitask learning in neural networks, emphasizing its applicability in various domains such as natural language processing, computer vision, and speech recognition. Recent advancements, including the use of attention mechanisms and meta-learning, have further improved the efficiency of multitask models.

MTL has emerged as a promising approach in advancing speech processing tasks by leveraging shared representations and learning efficiencies across multiple related tasks. In the context of speech assessment, MTL has played a crucial role in enhancing the accuracy and robustness of models tasked with evaluating spoken language proficiency. Early research in speech processing-related MTL focused on jointly learning acoustic modeling\cite{IntegratingFA}, phonetic classification\cite{Phone-Aware-MTL}, and language modeling tasks.\cite{weber-etal}, laying the foundation for subsequent advancements in speech assessment models. Building upon this foundation, recent studies have extended MTL techniques to encompass a broader range of speech assessment tasks, including speech intelligibility, speech quality, and analyzing other dimensions crucial for effective communication. By jointly modeling these diverse tasks,  researchers have demonstrated improvements in overall assessment accuracy\cite{mosa-net, mbi-net}. Despite these advancements, challenges remain in addressing domain-specific nuances and scaling MTL approaches to large-scale speech assessment datasets. Ongoing research efforts continue to explore novel architectures and learning paradigms to further enhance the effectiveness of MTL in the context of speech assessment. However, Ongoing research efforts continue to explore novel architectures and learning paradigms to further enhance the effectiveness of MTL. Moreover, the exploration of different speech assessments or related features, such as contextual information, speaker characteristics, and environmental factors, is underway to improve the performance of individual tasks within the multitask learning framework. 

In the context of acoustic scenes, utilising a multi-task DNN model has been seen to improve results in a soundscape task previously \cite{mitchell2023deep}, whereby simultaneously predicting noise annoyance alongside individual sound events, leads to an increase in accuracy of the subjective noise annoyance prediction. As well as showing the predictive power of multi-task learning, this supports the hypothesis that contextual information, alongside sonic characteristics, are important when considering the perceptual relationship we wish to have with our acoustic environments.

\vspace{1em}

% Include layer-wise diagram of model
\section{System Outline}

In this section, we introduce the fundamental parts of the proposed system (see Figure 1), including the SNR Control Mechanism dependent on the acoustic context, Acoustic Context Modelling and Prediction including feature representations via the attention mechanism and t-SNE dimensionality reduction.

\subsection{SNR Control Mechanism}

The basic idea here is to utilise the user preferences and the environmental SSNR prediction in order to control in real-time the target $SNR^*$ of the DNN based SE system. %The proportionality of $SNR^*$ to the preference function $f(\widehat{SNR}, A)$ arises because the synthesised output of the IRM is assumed to approach the true $SNR$ (by definition of the SE learning objective).
\begin{equation}
SNR^* \: \alpha \: f(A) \: and \: A =\widehat{SNR} \: \beta + \gamma
\end{equation}
\vspace{-0.2in}
\begin{equation}
\:  A \: \alpha \: (1-p) | \; \widehat{SNR} \:  where \: (0 \leq \;p_{1:N} \leq 1)
\end{equation}

where $\widehat{SNR}$ is the predicted environmental SNR. The elicitation sequence begins at 50\% enhancement and $(0 \leq A \leq 1)$ for input into the activation function (equation \ref{eq:log}). $p$ is the preference from the elicitation phase and $\beta$ and $\gamma$ are learnt from inputs $p$ and $\widehat{SNR}$. %and are dependent on acoustic scene index $c$. \\

For each acoustic scene $c$, we derive a personal preference function that is modelled from the elicitation phase. Then, we have the ability to predict the preference for $A$ based on the joint prediction of the SSNR ($\widehat{SNR}$) and the acoustic scene. In this case a closed set $\textbf{A}$ denoted by the acoustic scene index $c$ refers to the set of preferences for each scene.
\begin{equation}
\begin{aligned}
     A_{c}= \widehat{SNR}\cdot \beta _{c}+\gamma_{c} \: where \: \ A_c \in \textbf{A}_{1:C}
\end{aligned}
\end{equation}

\begin{equation}
SNR^* \: \alpha \: A_c + \frac{(K -  A_c)}{ (C + Qe^{-Bt})^1/v} 
\label{eq:log}
\end{equation}

In the \textbf{Generalised Logistic Activation Function (Richards Curve)}, the lower asymptote, denoted as $A$ (which serves as a noise floor within $SNR^*$), is to be determined based on the learned preference function. Equation \ref{eq:log} adjusts the activation and, consequently, the resultant $SNR$ for the ultimate output layer via the mask magnitude spectrogram estimation of the IRM-SE model. %This methodology is expected to extend to various mask types, as will be illustrated in future research.

\subsection{Acoustic Context Modelling}

\subsubsection{Multi-Task Deep Neural Network}

\begin{figure}
    \centering
    \includegraphics[scale=0.4]{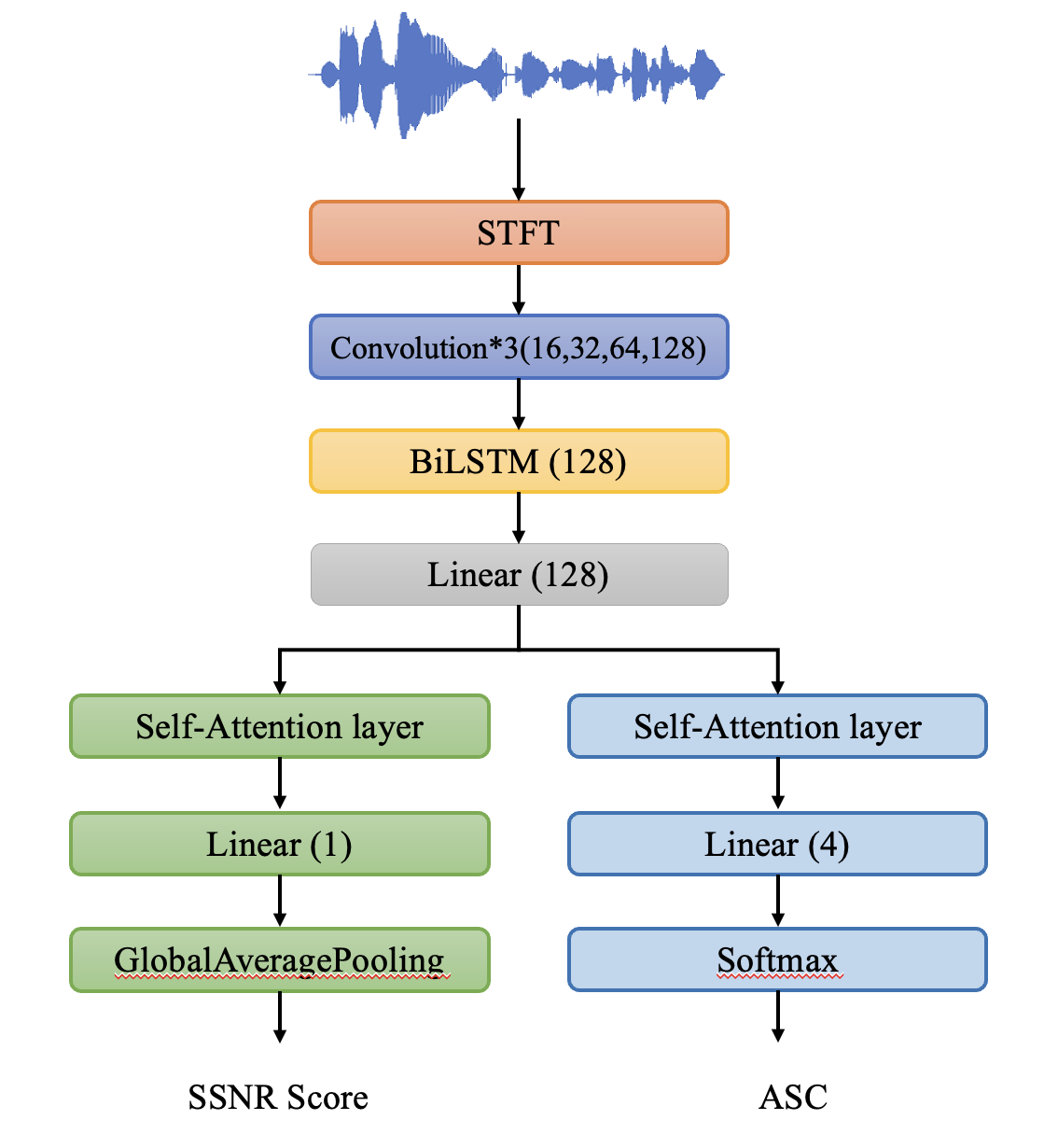}
    \caption{The Multi-Task Acoustic Scene Classification and SNR Prediction Model illustrating a shared encoded feature space.} 
    \label{fig:multi-task_model_overview}
\end{figure}

The SNR and Acoustic scene are predicted in a multi-task model according to the architecture in Figure 2. We have chosen power spectral features as input and CNN-BiLSTM with self-attention as a model architecture after experimentation. Our model architecture is composed of 4 convolutional blocks, each comprising 3 convolutional layers. The number of filters in each block follows an incremental style, with 16, 32, 64, and 128 filters, respectively. These convolutional blocks play a crucial role in extracting high-level signal features from the input data. Following the convolutional layers, a single-layered BiLSTM with 128 units is employed. The BiLSTM is designed to capture the sequential dependencies within the extracted features. This is followed by a fully connected layer with 128 units. The resulting encoded feature space is shared between two branches, each dedicated to specifically predicting SNR or ASC. Both branches incorporate an attention layer to enhance the model's focus on relevant information.

For SNR prediction, an additional time-distributed linear layer with a single unit is introduced after the attention layer. This facilitates the frame-wise prediction of SNR scores. To calculate the utterance score, Global Average Pooling is applied to the frame-wise scores. For ASC, a linear layer with 4 units is employed to predict the corresponding labels. This comprehensive architecture allows our model to effectively capture and utilise intricate features for the multitask prediction of SNR and ASC, leveraging the strengths of convolutional, recurrent, and attention mechanisms.

Because of the shared convolutional deep feature encoding we are able to also save significant computational cost. If we were to have two single task models, the cost would be increased whilst distributed between multiple feature encodings. Apart from performance increase this is another benefit of the multi-task framework.
%The CNN-BiLSTM+AT model architecture has 3 convolution blocks each consisting of 4 convolutional layers (16, 32, 64, and 128 filters), followed by single-layered BiLSTM (128 units), a fully connected layer (128 units). This encoded feature space is then shared amonst the SSNR and ASC prediction models.

\subsubsection{Environmental Segmental Signal-to-Noise Ratio Estimation}

Signal-to-noise ratio (SNR) is a basic objective speech quality measure which quantifies the relative amount of background noise present in a speech clip, usually in terms of sound pressure level (SPL), measured in decibels (dBs). It is defined as the ratio of signal intensity to noise intensity. In contrast to working directly on the entire signal in our experiments, we used the Segmental Signal-to-Noise Ratio (SNRseg or SSNR) \cite{segsnr}, which calculates the average of the SNR values (in dB) of short segments (15 to 20 ms) given as:

% \vspace{-0.3in}
% %\begin{figure}
% \begin{center}
% \begin{equation}
% \includegraphics[width=0.5\linewidth]{emsnewsletter-template/SSNR.png} \label{eq:seg_snr}
% \end{equation}
% \end{center}

\begin{equation}
\label{seg_snr}
\text{Seg.SNR} = \frac{10}{M} \sum_{m=0}^{M-1} \log_{10}\left( \frac{\sum_{n=N_m}^{N_m+N-1} s^2(n)}{\sum_{n=N_m}^{N_m+N-1} [s(n) - \hat{s}(n)]^2}\right)
\end{equation}

%\caption{SSNR prediction model}
%\end{figure}
\textit{M} represents the number of segments, \textit{s(n)} represents the clean signal sample at time \textit{n}, \(\hat{s}(n)\) corresponds to the sample of the processed (noisy or enhanced) signal at the time \textit{n}, and \textit{N} denotes the total number of samples in the segment. 

The user's preferred SNR essentially refers to the SNR that a user prefers when listening to a target speaker, whilst experiencing a relative SPL of background noise.  For a non-intrusive measure of SSNR, we have built a SSNR prediction model for our proposed framework.% The output of the attention layer is fed to a fully connected layer (1 unit) a global average operation was then used to produce the prediction score as illustrated in Fig. ~\ref{fig:ssnr}.

%\begin{figure}[!t]
%    %\centering
%    \begin{center}
%    \includegraphics[width = 0.5\columnwidth]{Screenshot 2023-03-23 at 20.48.32.png}
%    \caption{SSNR prediction model}
%    \label{fig:ssnr}
%    \end{center}
%\end{figure}
\vspace{-5mm}
\subsection{Evaluation}
To evaluate the SSNR prediction model, three evaluation metrics were used: linear correlation coefficient (LCC), Spearman rank correlation coefficient (SRCC) and mean squared error (MSE) \cite{stoinet}. Higher LCC and SRCC scores show that the predicted scores are of higher correlations to the ground truth assessment scores, whilst a lower MSE score indicates that the predicted scores are closer to the ground-truth assessment scores. The experimental results of the SSNR prediction model using cross-validation for the GRID-CHIME3 corpus are shown in table 1. 

\begin{figure}
    \centering
    \includegraphics[scale=0.5]{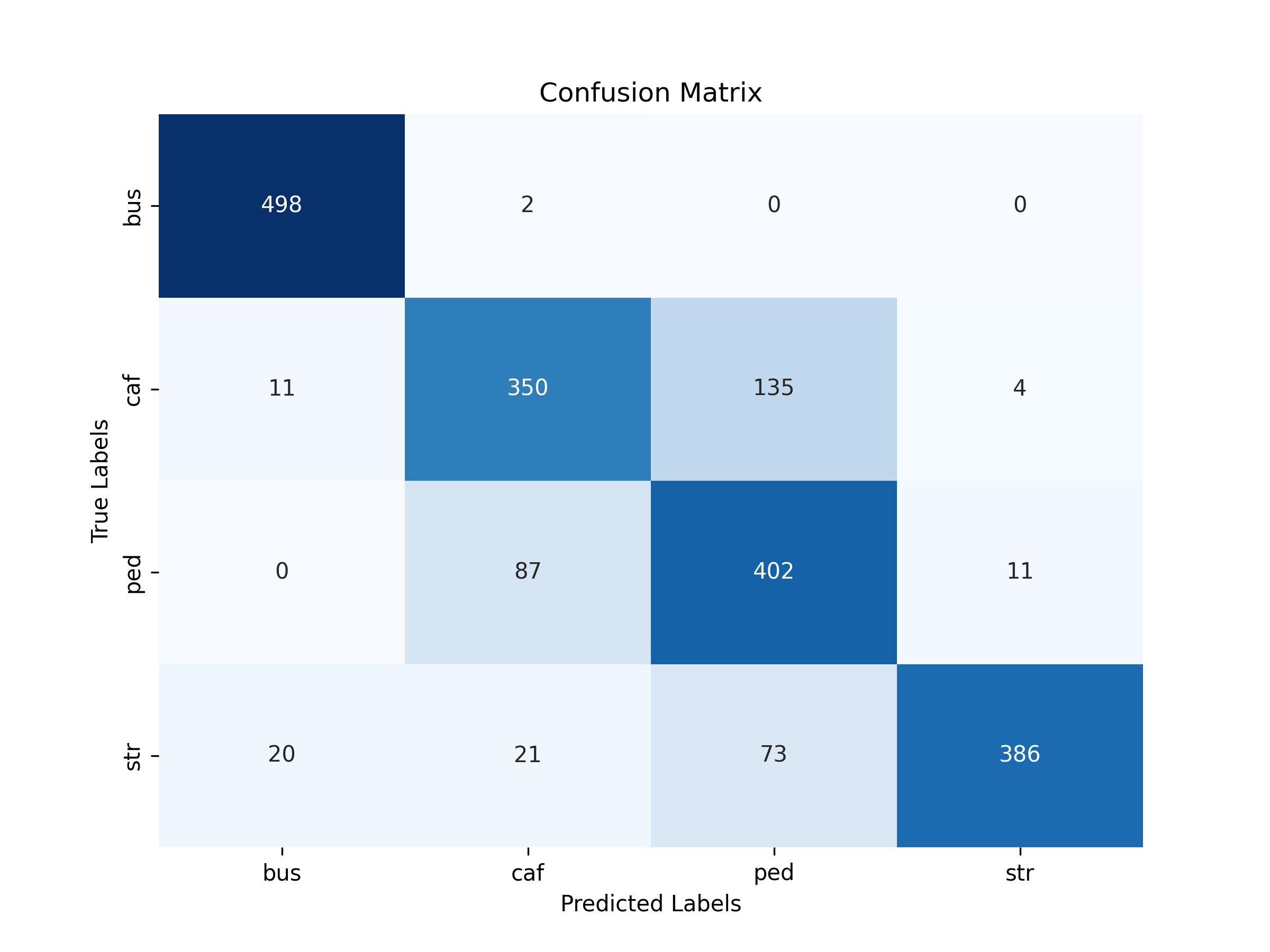}
    \includegraphics[scale=0.5]{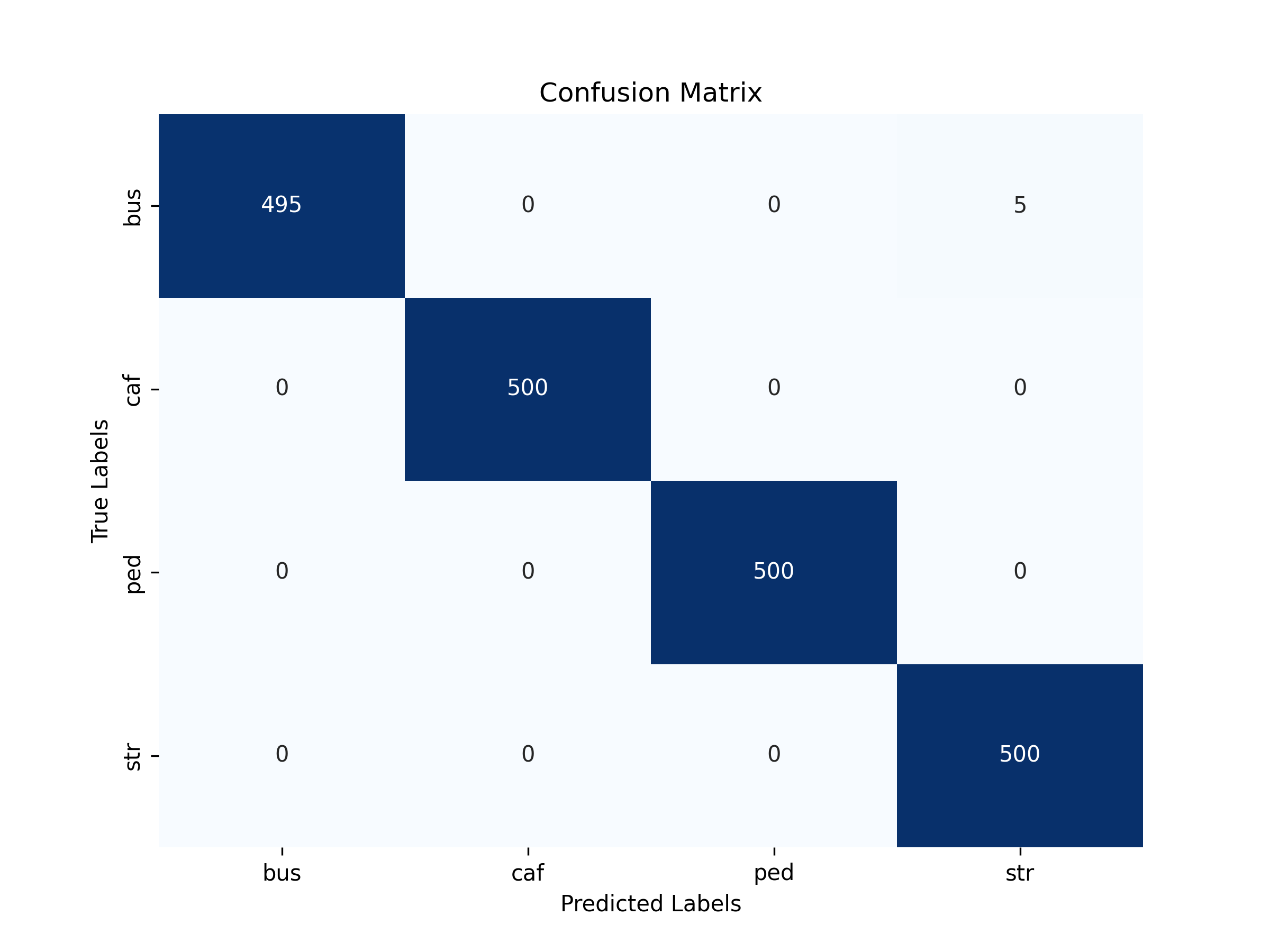}
    \caption{Confusion Matrices for the Single-task (above) and Multi-task Model (below). The multi-task achieves near perfect prediction performance on the held out set whilst the single task model confuses the acoustic scenes without clear long droning background sounds such as the bus engine. The pedestrian and cafe scenes are often confused in the single-task model which in Figure 4 there is no clear separable boundary between classes in feature representation, though this is improved in the multi-task model, displayed in Figure 5.}
    \label{fig:confus-matr}
\end{figure}

Figure \ref{fig:confus-matr} shows the classification scores for the Acoustic Scenes in the GRID-CHIME3 dataset.

\begin{table}[]
\label{tab:snr}
\centering
\begin{tabular}{lllll}
\cellcolor[HTML]{4472C4}{\color[HTML]{FFFFFF} \textbf{Model Type}} & \cellcolor[HTML]{4472C4}{\color[HTML]{FFFFFF} \textbf{LCC}} & \cellcolor[HTML]{4472C4}{\color[HTML]{FFFFFF} \textbf{SRCC}} & \cellcolor[HTML]{4472C4}{\color[HTML]{FFFFFF} \textbf{MSE}} &  \\
\cellcolor[HTML]{CDD4EA}Single-Task & \cellcolor[HTML]{CDD4EA}0.979                              & \cellcolor[HTML]{CDD4EA}0.969                               & \cellcolor[HTML]{CDD4EA}0.393                              &  \\
\cellcolor[HTML]{E8EBF5}Multi-Task  & \cellcolor[HTML]{E8EBF5}0.981                              & \cellcolor[HTML]{E8EBF5}0.970                               & \cellcolor[HTML]{E8EBF5}0.343                              &  \\
                                    &                                                             &                                                              &                                                             & 
\end{tabular}
\caption{SSNR Scores on the GRID-CHIME3 dataset}
\end{table}
%   
% LCC
% SRCC
% MSE
% Single-Task
% 0.9790
% 0.9689
% 0.3933
% Multi-Task
% 0.9813
% 0.9698
% 0.3434
%

\subsection{Encoded Acoustic Scene Feature Representation}

\subsubsection{Dimensionality Reduction}

\begin{figure}
\centering
\includegraphics[scale=0.15]{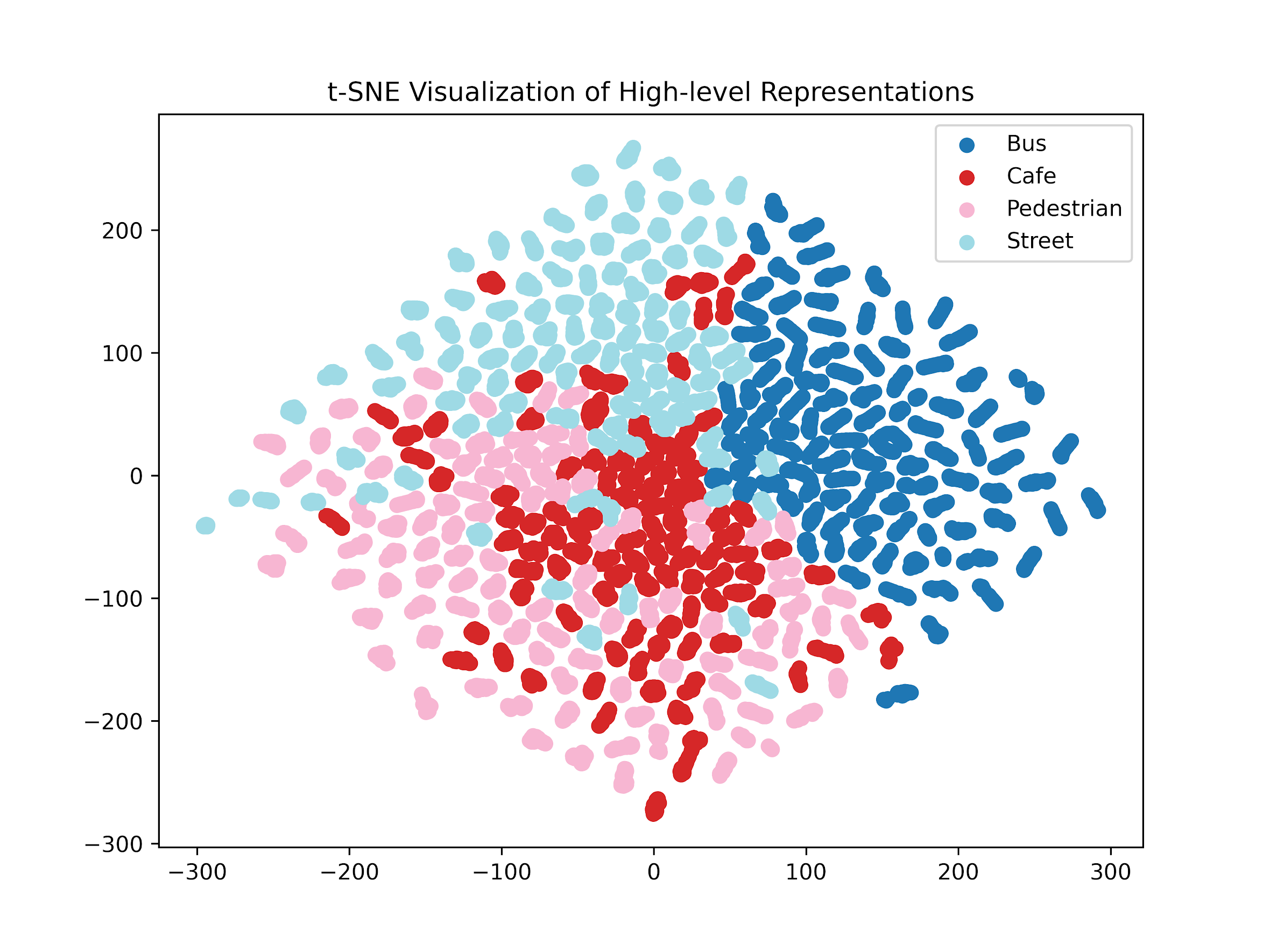}
\caption{T-SNE Visualisation of Single-task ASC feature embeddings at the self-attention layer level. There is a lot of overlap in the feature space between the Cafe, Pedestrian and Street scenes.}
\label{fig:tsne1}
\end{figure}

\begin{figure}
    \centering
    \includegraphics[scale=0.6]{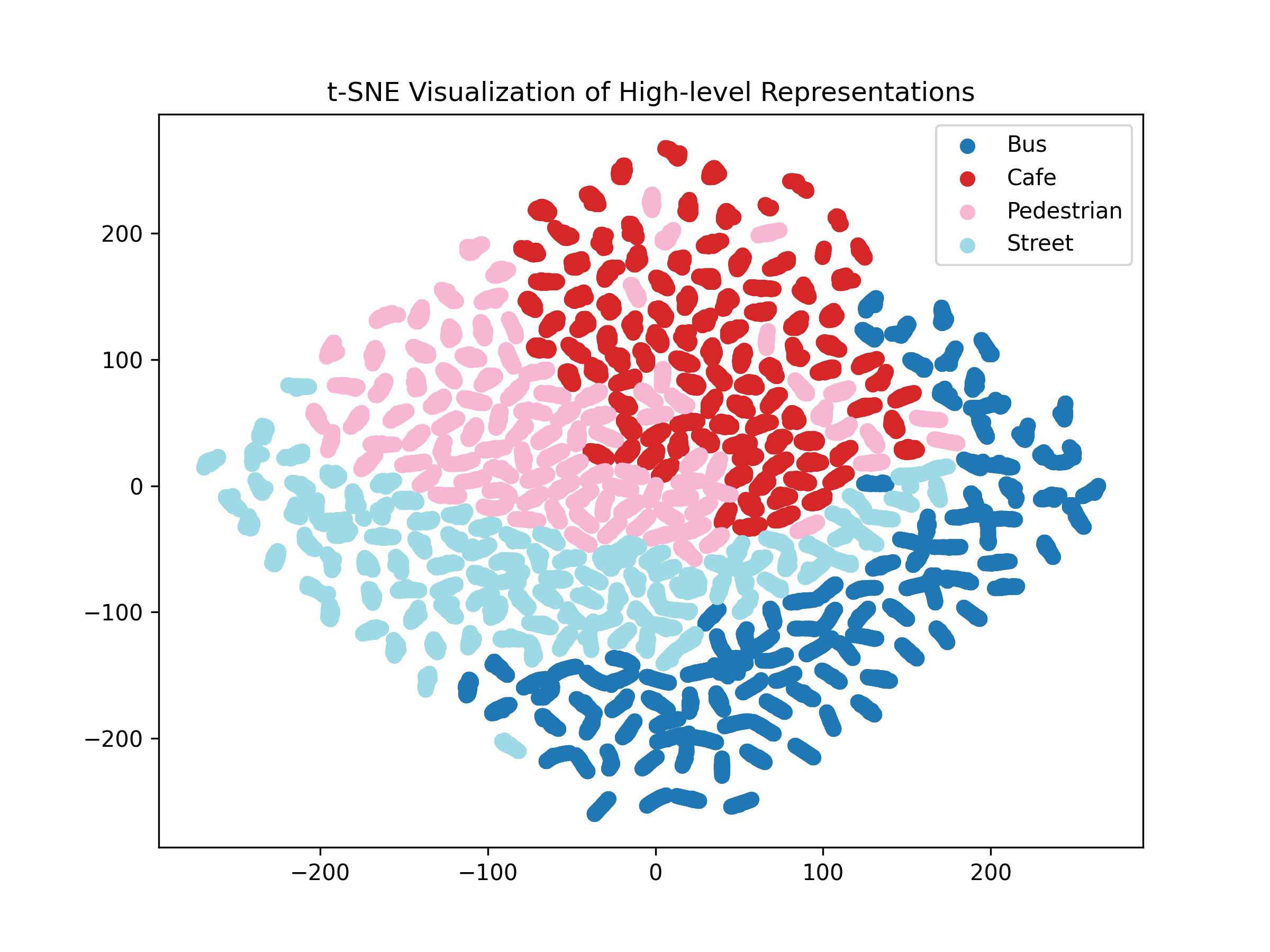}
    \caption{T-SNE Visualisation of Multi-task ASC feature embeddings at the self-attention layer level. The representation is much more well defined than the single-task model, though the classes are not completely separable.}
    \label{fig:tsne2}
\end{figure}

\begin{figure}
    \centering
    \includegraphics[scale=0.6]{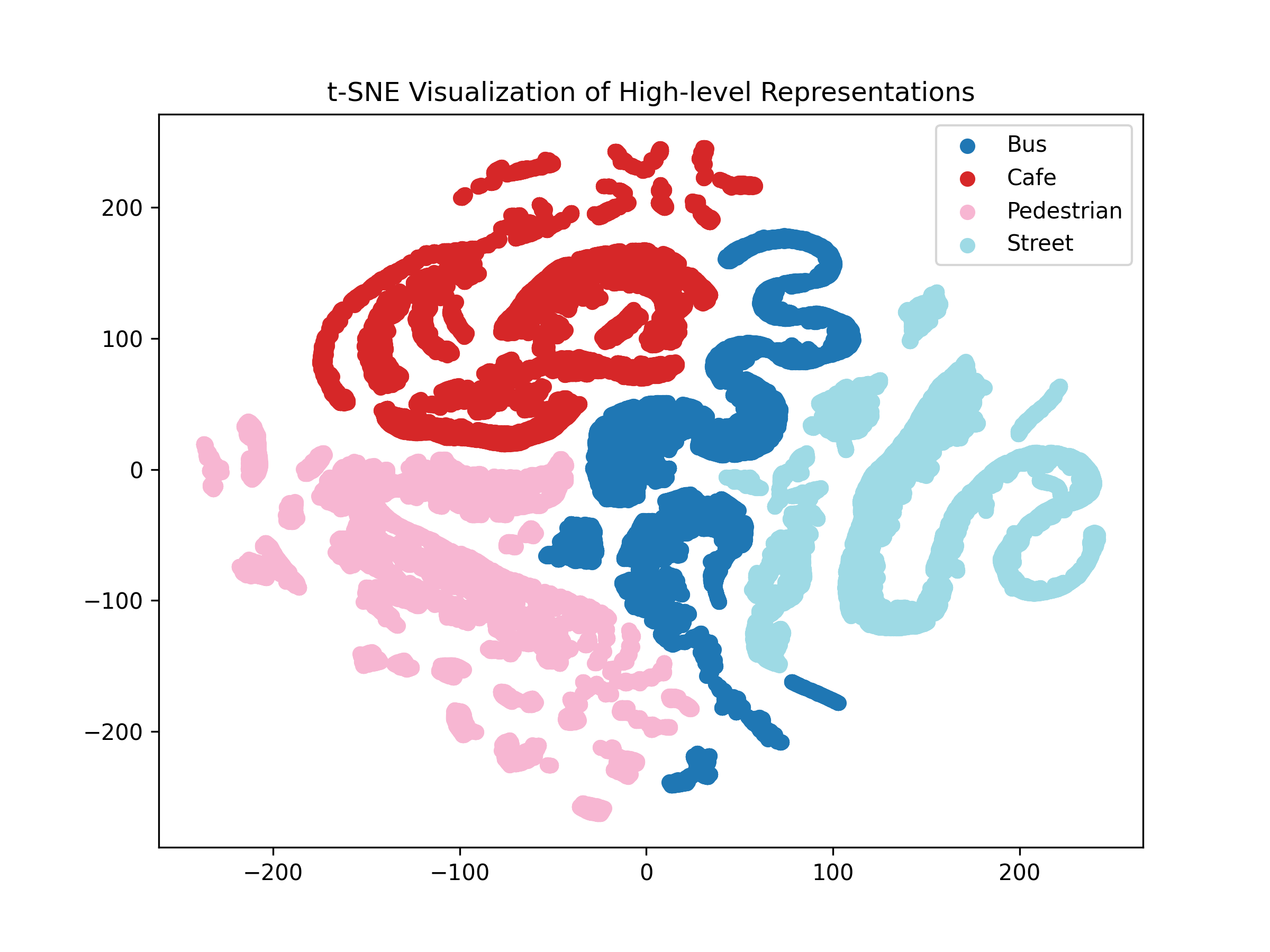}
    \caption{T-SNE Visualisation of Multi-task ASC feature embeddings at the final Linear (4 neuron) layer. The classes now show distinct boundaries according to just two dimensions visualised here. }
    \label{fig:tsne3}
\end{figure}

% Through dimensionality reduction technique t-SNE (cite) on the encoded Neural Network features from the attention layer \ref{fig:tsne1}, and the Dense layer \ref{fig:tsne2} (Linear 128?), we can achieve the following visualisations, which are also interpretable feature spaces which can be utilised for modelling purposes, and potentially clinical use when discussing HA outcomes.
Through the application of dimensionality reduction technique t-SNE we can gain some interoperability on the encoded DNN features. We also present the final linear layer of the multi-task framework (Figure \ref{fig:tsne3}) to show the resultant linear separability between classes. t-SNE, is a powerful method for visualising high-dimensional data in a lower-dimensional space while preserving local structures \cite{tsne_info}. In our analysis, t-SNE was configured with two components, a perplexity value of 50, and 5000 iterations. These settings enable the creation of visualisations that offer meaningful insights into feature spaces. We present the following visualisations for the ASC task from the self-attention layer: in Figure \ref{fig:tsne1} we have the single-task model, and in Figure \ref{fig:tsne2}, the multi-task model. Such visualisations not only aid in modelling tasks, helping to address problems such as cold-start and generalisation capability once the model scales in the real-world, but also hold potential relevance in clinical discussions, particularly concerning Hearing Aid (HA) outcomes with respect to different environments. % link this in future work

\subsubsection{Attention Layer Representation}

\begin{figure}
    \centering
    \hspace{0.27cm}\includegraphics[scale=0.177]{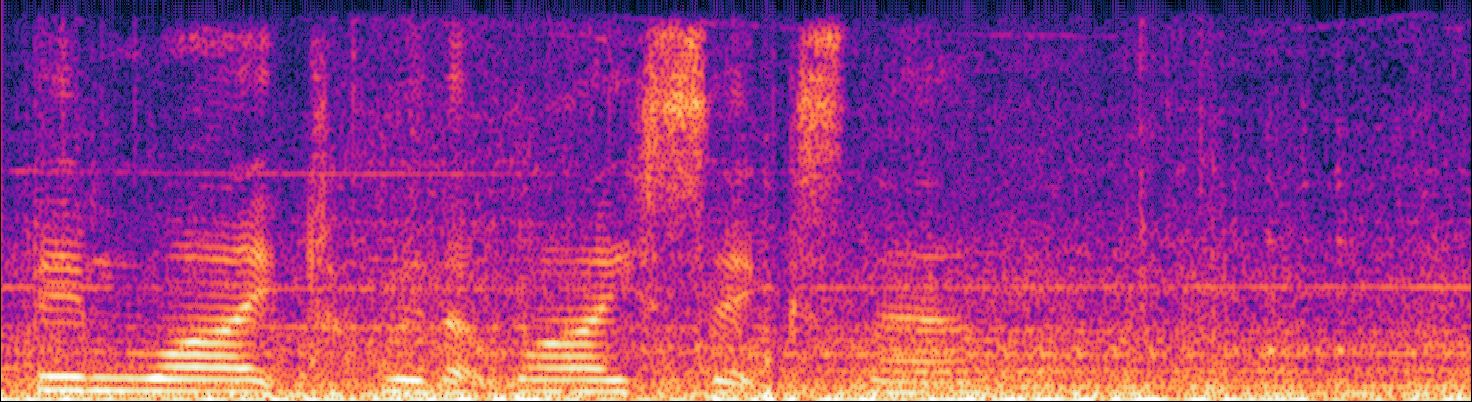} \\
    %\caption{The Multi-Task Acoustic Scene and SNR Prediction Model} 
    \includegraphics[scale=0.35]{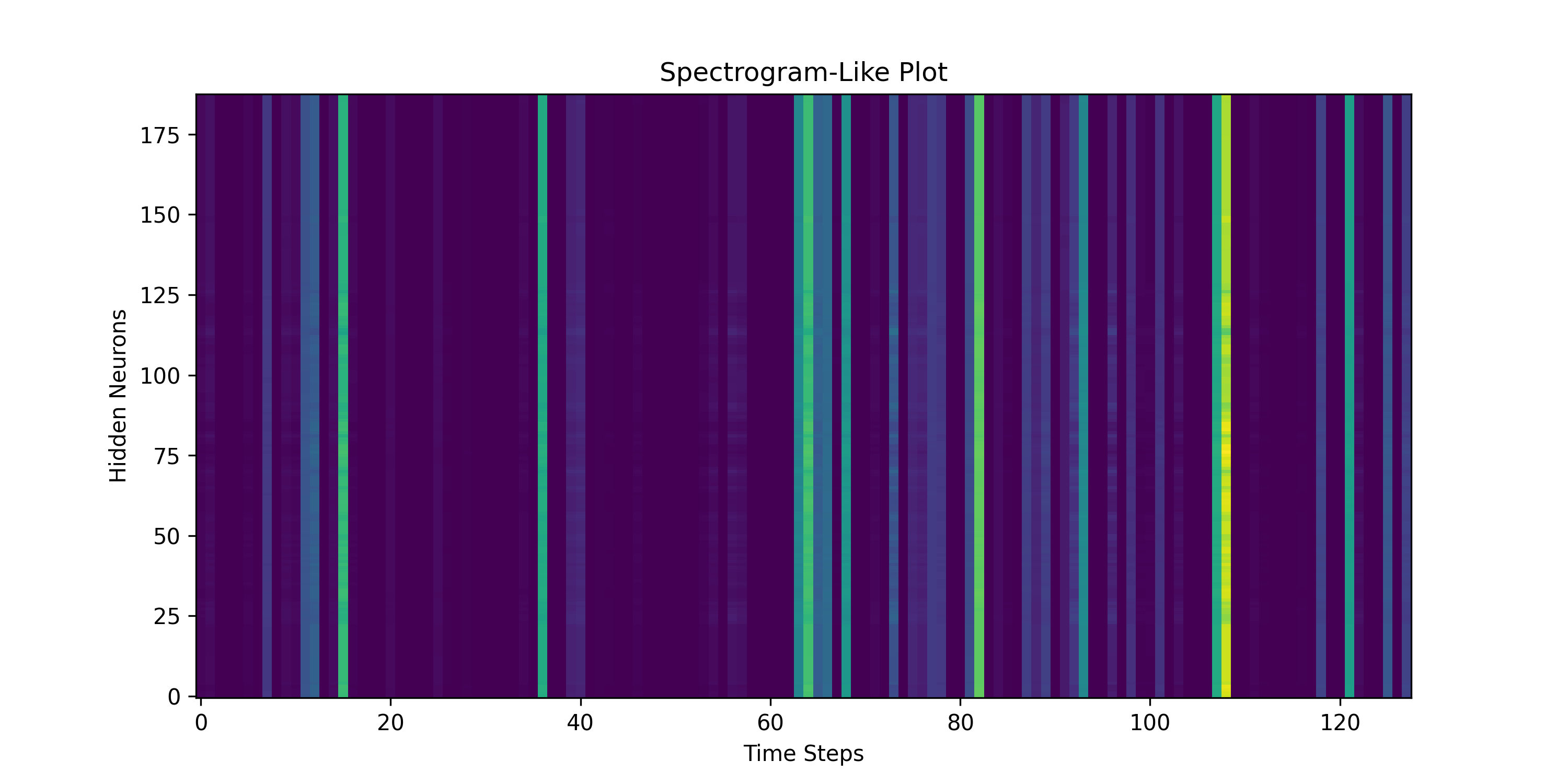}
    \includegraphics[scale=0.35]{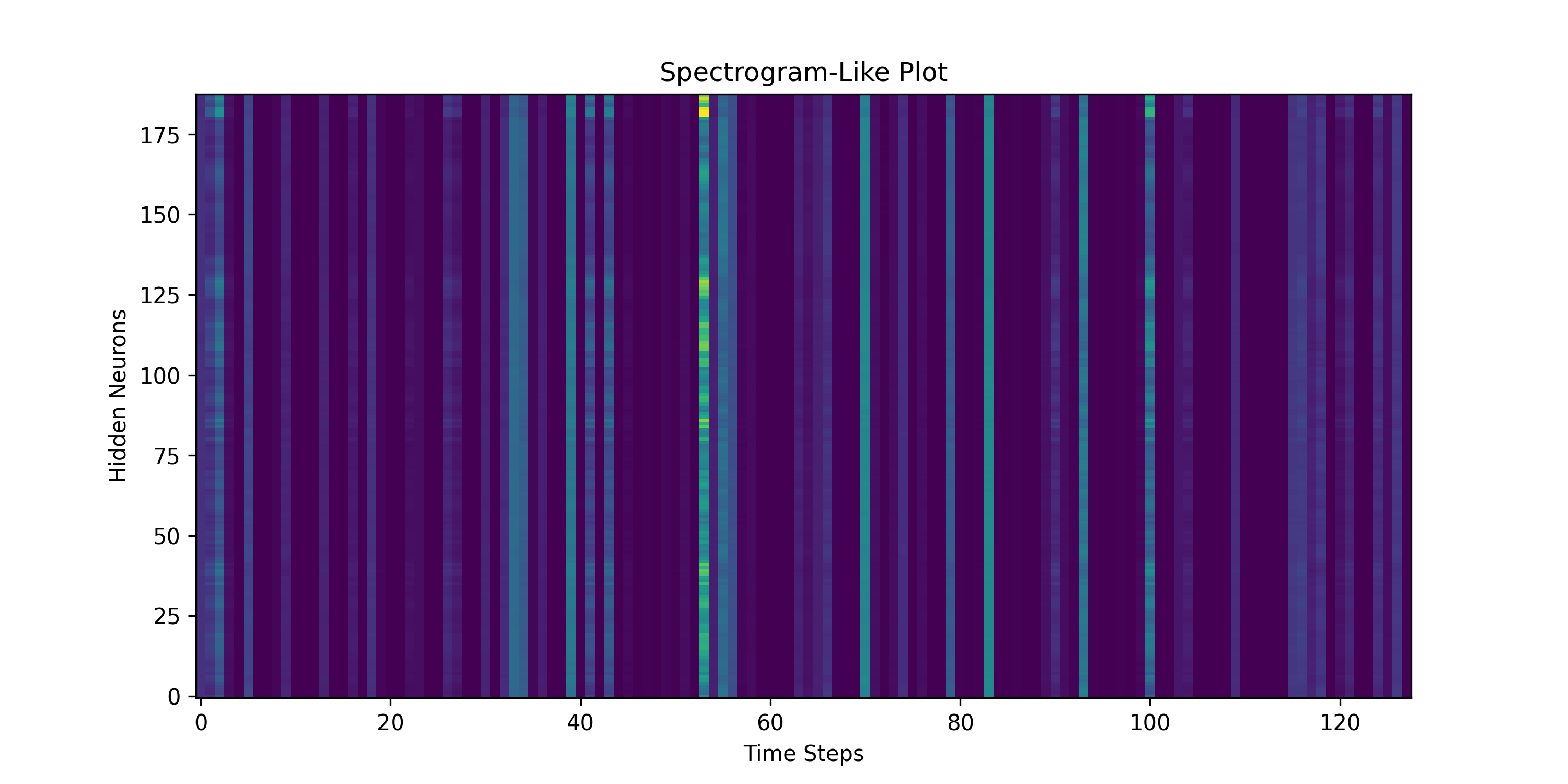}
    \caption{From top to bottom (bus noise type): the raw noisy spectogram; single-task ASC and multi-task ASC prediction. The multi-task model is able to more selectively activate neurons across the spectro-temporal range to distinguish salient sound events that aid classification (such as the buses' distinctive hum, which can be seen across the time steps here).} 
    \label{fig:attention-representation}
\end{figure}

In Figure \ref{fig:attention-representation}, the mechanics of the model can be loosely interpreted with respect to the raw spectogram of the noisy mixture. Improvement in performance of the model can be explained by more granular neuronal activation, indicating more specialised neurons attending different spectro-temporal patterns, and more non-speech time-distributed neuronal activation for the ASC. 

The SSNR prediction also sees a slight improvement in performance with the multi-task model, and thus the attention representation can be seen to be more concentrated around speech regions (plot omitted for brevity).

% mark huckvales new corpus would allow more than 2 target speaker switching in VR, however, it is not multi-sensory (RF & thermal imaging), and does not take place in a real environment to allow for ecologically valid acoustic complexities and lombard speech

\section{Experimental Setup} \label{sec:setup}

The subjective study was approved by the Ethics Committee of Napier University, Edinburgh on 27 November 2023.

\subsection{Presentation}

Users interact with the multi-environment demo on a laptop, tuning their preferences in real-time to GRID-CHIME3 data. The data used, $\mathbf{Y}$, for this study is the GRID-CHIME3, augmented dataset ~\cite{cooke2006audio, chime-3}. For each of the experimental phases, one unique sentence is played from one of 4 target speakers (2 male, 2 female) from the GRID corpus are overlaid onto each of 5 noise levels sampled from the 4 different environmental background noises of the CHIME3 dataset, cut and volume adjusted. This leads to 5 SNRs: (-9,3,0,3,9) dB.

%then new data will be presented (recorded with microphone in similar environments around Edinburgh; utilising the ASPIRE dataset), 
%The user will be able to see how the preference generalises across acoustic scenes. 

\noindent The experiments are split up into a preference elicitation stage and an evaluation stage. In the Preference Elicitation Stage, the participant engages with the `up, down or no change' interface. The evaluation stage is comprised of 3 conditions:\\ 
\textbf{1. Noisy Condition:} Utterances at each of the 5 SNR's played in each of the 4 simulated environments.\\
\textbf{2. MaxSE Evaluation Condition}: SE with maximum $SNR^*$ , i.e. sigmoid-scaled final output layer for the IRM-SE.\\
\textbf{3. PLSE Evaluation Condition}: SE with the inferred $SNR^*$ from the learned preference function, i.e. output layer scaled according to equation 1.

All of the experiments were in rooms with below 35dB ambient noise according to an audiometer. 

\subsection{Normal Hearing vs Hearing Impaired Playback}

The HI participants were all fitted with HA's according to their prescription with NAL-NL2 procedure, the sound was then played back directly through the hearing aids via Bluetooth at a comfortable volume for each participant (set before the start of the experiment). The NH participants were played the sound through a pair of Sony WH-1000XM4 on-ear headphones and the sound and video were played back via a Macbook Pro laptop. %The video was displayed on a 16-inch M1 Max Macbook Pro.

\subsection{Ambient Occlusion}
\cite{ear-occluded-intel} shows that intelligibility improves greatly when there is more sound occluded via the ear canal. New materials have made it possible to achieve a lot of attenuation simply with ear buds. Additionally, recent technology has provided excellent active noise cancelling abilities that are widely available on the market.%via the use of standing waves %(cite?). 
These advancements in small acoustical technologies have made it possible to remove the vast amount of ambient sound from entering the ear canal directly, in earbuds, wearables or potentially hearing aids, especially for those who have a hearing loss to begin with.

For these reasons, the experimental set up assumes an asymptotic, complete ambient occlusion via direct streaming of noisy audio into the hearing aid. Future studies may look at the extra variablity different acoustic couplings may introduce (e.g. by introducing free-field background sound in-situ alongside the PLSE).

%ALTERATION:
%All future experiments will be made with an additional 4th phase (No enhancement) to test the efficacy of the enhancement algorithm - without preference learning - with the individual).

\begin{figure}
    \centering
    \includegraphics[scale=0.7]{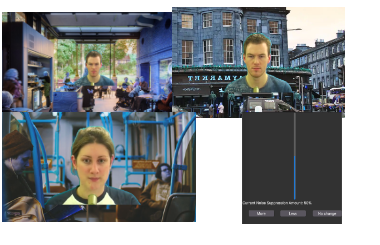}
    \caption{Preference Elicitation Phase with 4 Acoustic Scenes from GRID-CHIME3 Dataset with Noise Suppression Control}
    \label{fig:enter-label}
\end{figure}

\section{Subjective Testing Results}

\subsection{Normal Hearing}

2 participants that were clinically reported within normal hearing thresholds and did not report any abnormal difficulty hearing speech in noise were employed for the study. They underwent the experimental procedure described in Section 4.

\subsubsection{NH Elicitation}
Initial Normal Hearing results show that preference for target AVSE SNR indeed differs considerably across various environments; additionally, it can be shown that these preference functions vary for different listeners. This is illustrated in Figure \ref{fig:multi-env-pref-plots-nh} which show the resultant prefrence functions from the elicitation stage. Participant one prefers maximum NR/SE in the cafe at high SNR's, and at low SNRs in the street area. However, when the SNR is high in the street area, the participant prefers no SE, this may be to preserve the naturalness of the signal when the intelligibility of the signal is high (a similar pattern is shown for the pedestrian area). In the mean function, the participant shows preference for strong SE whilst maintaining some of the original signal in the audio.

The preference for participant two follows similar patterns, however, this participant shows much more disdain for the SE in general, with the preference functions not showing maximum SE in any circumstance. The mean function shows preference for some SE, but a dislike for it in circumstances like a quiet street or pedestrian area.

\subsubsection{NH Evaluation}

\begin{figure}
    \centering
\includegraphics[width=0.9\linewidth]{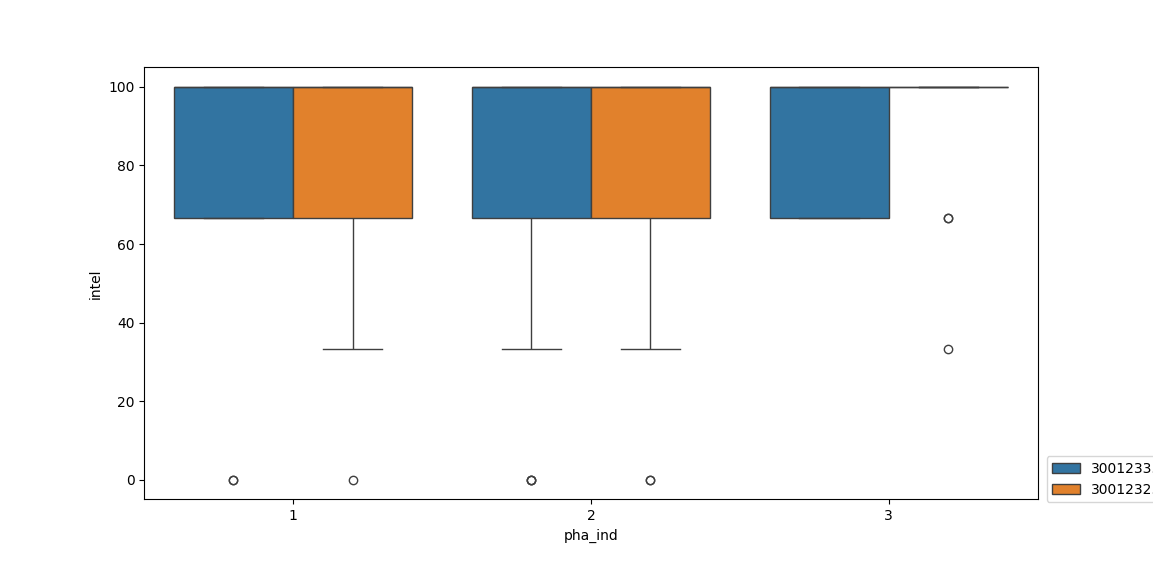}
\includegraphics[width=0.9\linewidth]{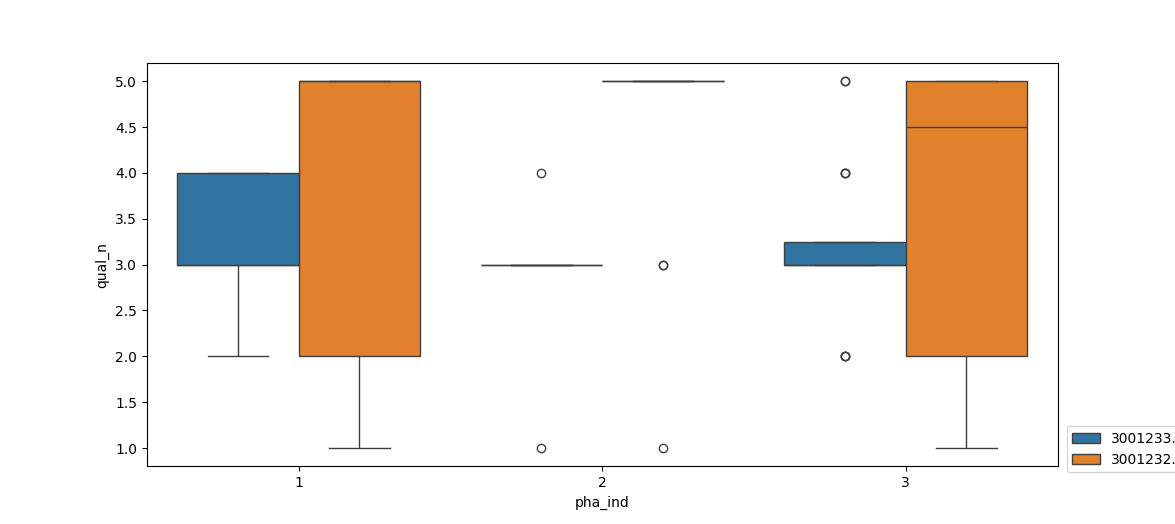}
\includegraphics[width=0.9\linewidth]{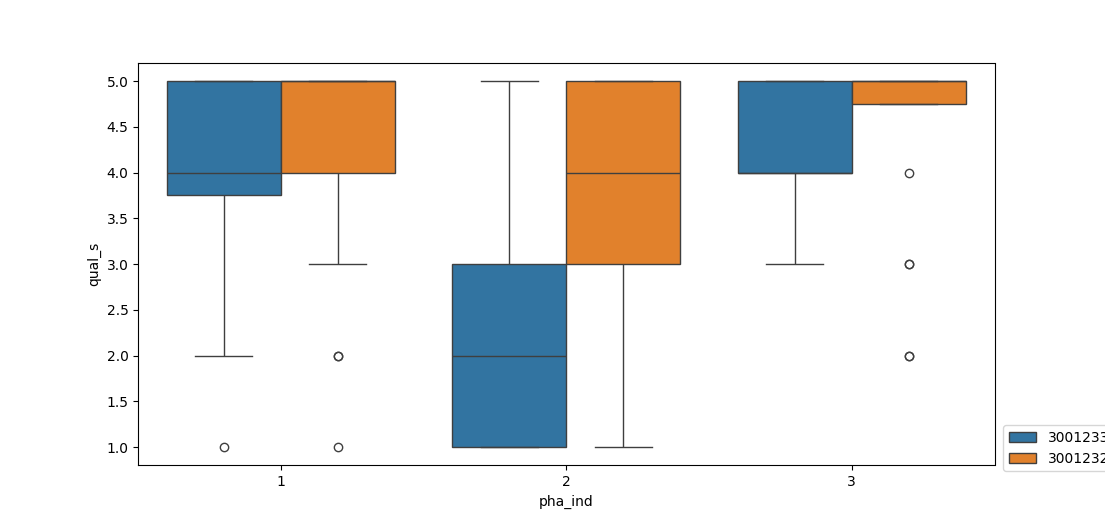}
    \caption{Normal Hearing Subjective Evaluation Results}
    \label{fig:NH_sub_res}
\end{figure}

Figure \ref{fig:NH_sub_res} shows Normal Hearing results from the evaluation stage of the testing. It can be seen that for NH-1, there is an improvement in speech quality as reported by the likert scale for the PLSE condition (C3). Similarly, this can be seen by NH-2. NH-2 interestingly also shows an improvement in intelligibility score in C3 as compared to C2; it can also be seen that the intelligibility score worsens for C2 (MaxSE) vs C1 for NH-1.

These findings indicate that the PLSE may be the best condition for NH listeners, despite the quality of the noise being apparently preferred for MaxSE for NH-2. In general in many of these situations, it would be unsafe and non-ideal to completely irradicate background noise, so it is thought that this slight case-wise exception to supporting PLSE in noise quality is due to a limitation of the study.

%\begin{table}[hspace]
%\begin{tabular}{llll}
%Participant   & Speech Quality (mean \pm std)        & Noise Quality                        & Intelligibility    \\
%NH-1 & 3.9 \pm 1.1 | 2.3 \pm 1.1 | 4.3 \pm 0.7   & 3.1 +- 0.7    & x  \\
%NH-2 & 2.3 \pm 1.1 x  & x & x \\
%NH-3 & x    & x     & x \\
%x &                                       &                          %            &                   
%\end{tabular}
%\caption{\label{hi_results}Normal Hearing Results, * indicates significance $p<0.05$ and ** indicates $p < 0.01$}
%\end{table}

\subsection{Hearing Impaired}

3 participants that were clinically reported as so:
HI-1: mild, HI-2: mild-moderate, HI-3: severe-profound. The experimental procedure was designed for the HI participants as per Section 4.

\subsubsection{HI Elicitation}

Figure \ref{fig:multi-env-pref-plots-hi} shows the elicitation results from the HI participants, in general it can be seen that the preferences differ greatly between each environment and each participant.
  %A video will be shown of the experimental set up.


\subsubsection{HI Evaluation}

\begin{figure}
    \centering
   \includegraphics[width=0.9\linewidth]{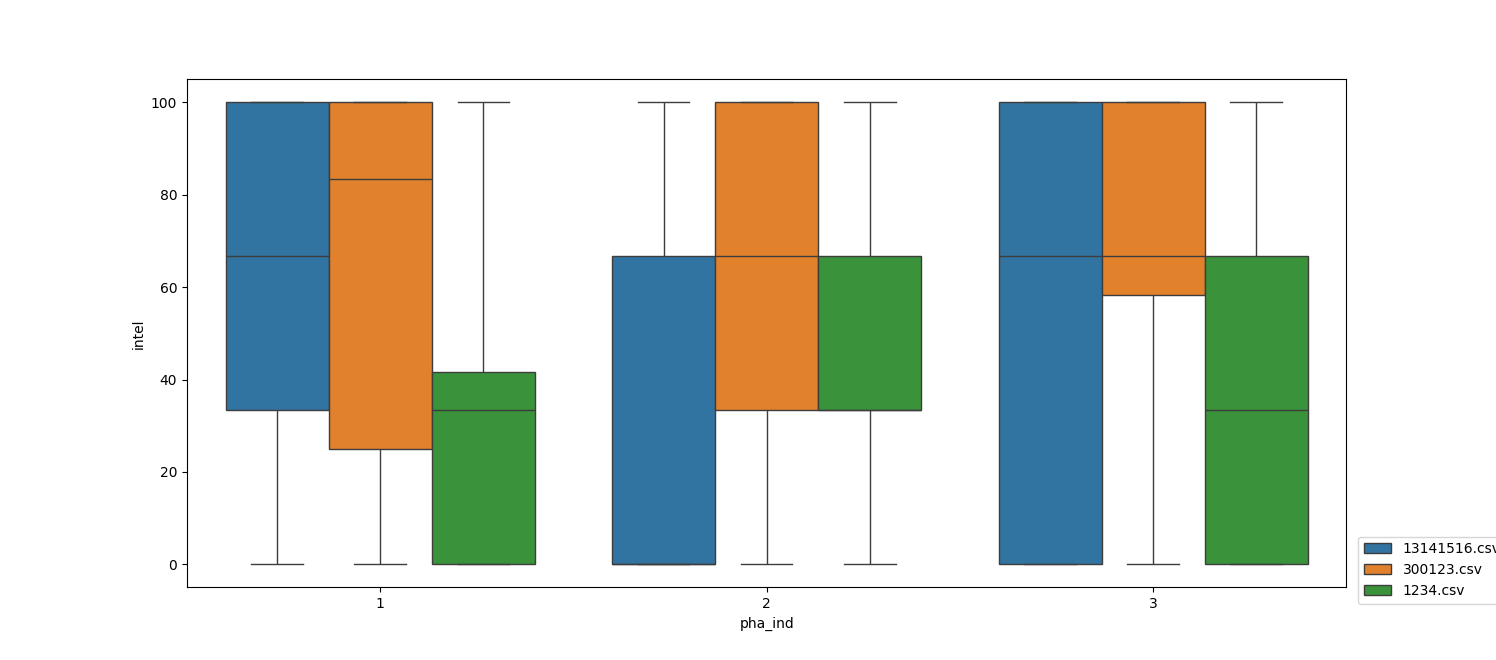} 
\includegraphics[width=0.9\linewidth]{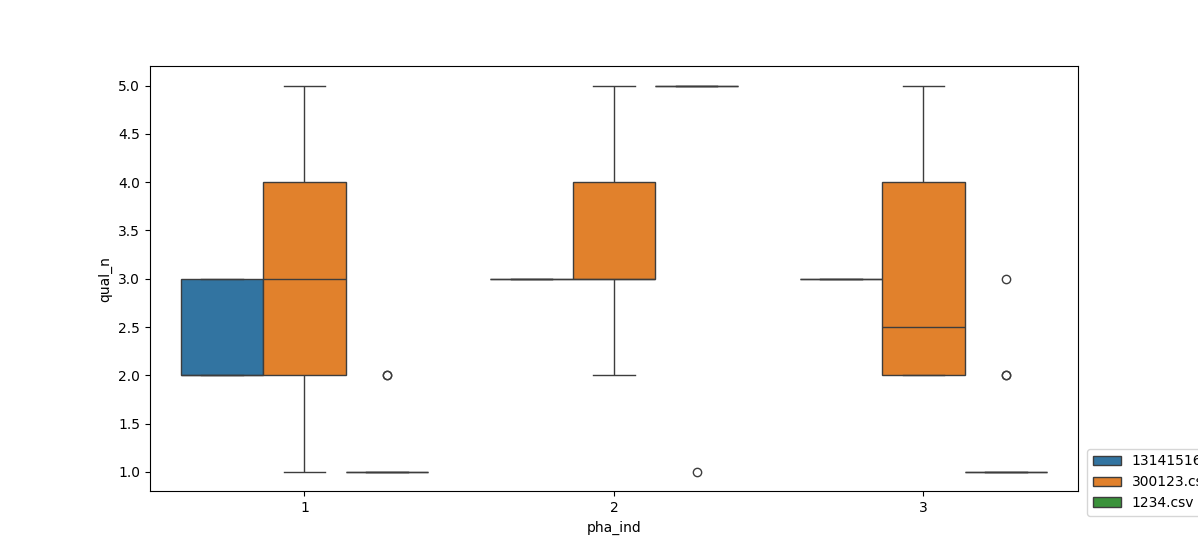}
\includegraphics[width=0.9\linewidth]{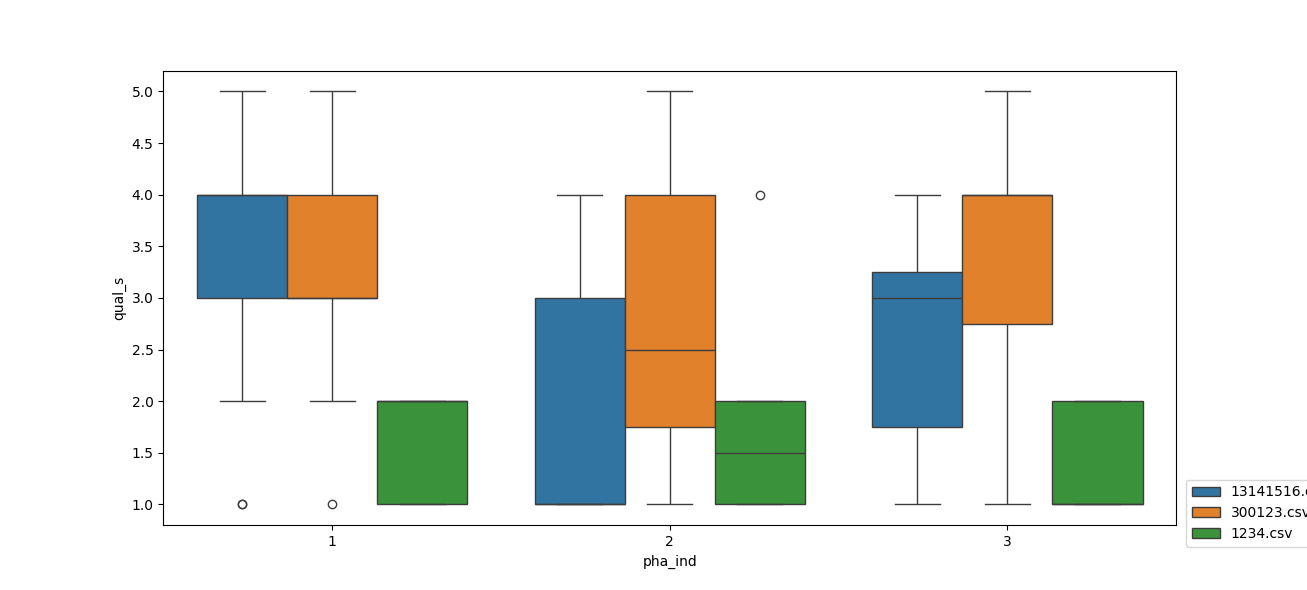}
    \caption{Hearing Impaired Subjective Evaluation Results}
    \label{fig:hi_se_res}
\end{figure}

\begin{figure}
    \centering
    \includegraphics[scale=0.5]{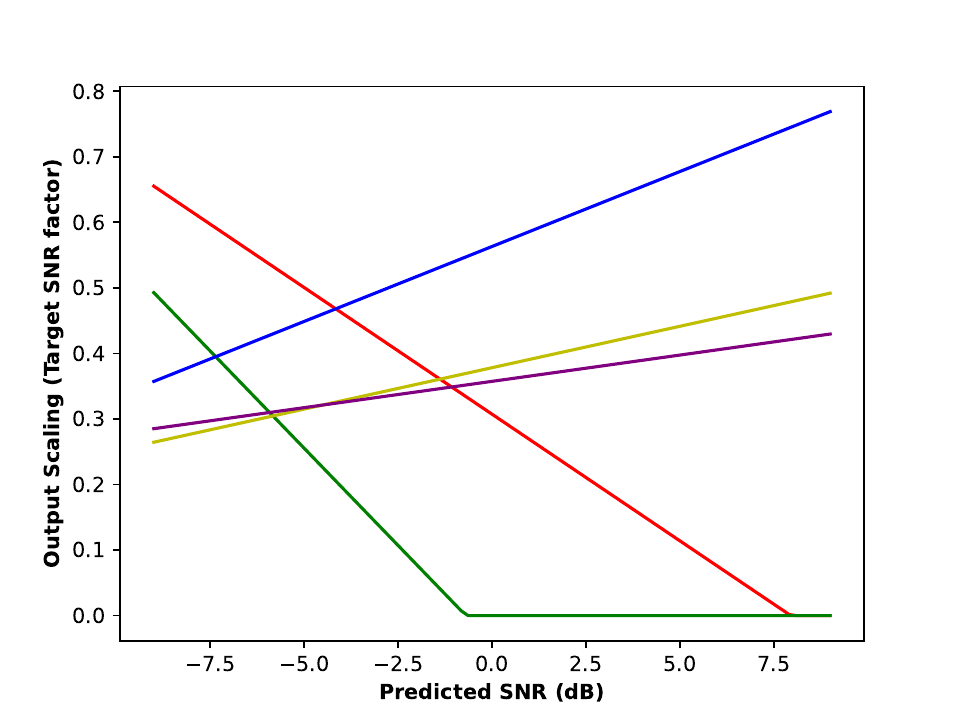} %why is this a png
    \includegraphics[scale=0.5]{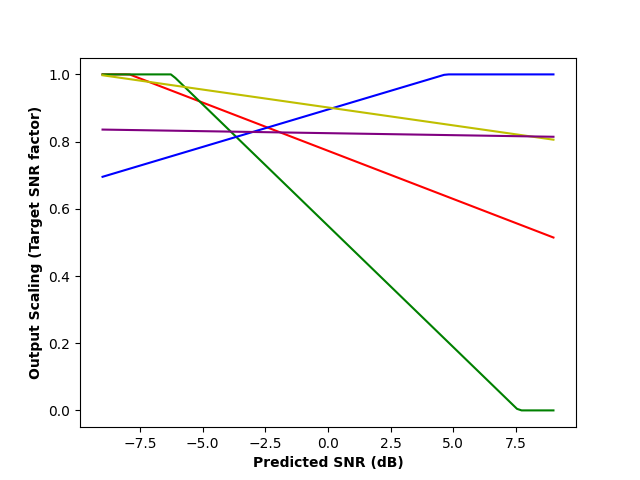}
    \caption{From top to bottom: Participant One, and Participant Two's Preference Functions in the CHiME-3 Acoustic Environments. Key: Pedestrian-Red; Street-Green, Cafe-Blue, Bus-Yellow; Mean Function-Purple} 
    \label{fig:multi-env-pref-plots-nh}
\end{figure}

\begin{figure}
    \centering
     %why is this a png
    %\includegraphics[scale=0.5]{5678.pdf}
    %\includegraphics[scale=0.5]{9101112.pdf}
    \includegraphics[scale=0.5]{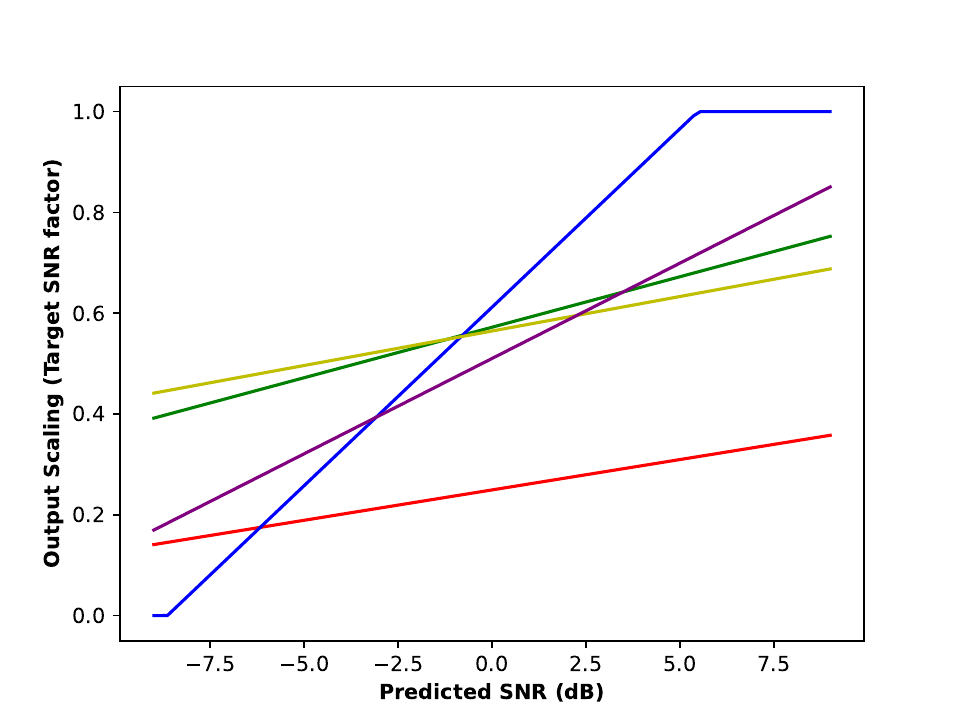}
    \includegraphics[scale=0.5]{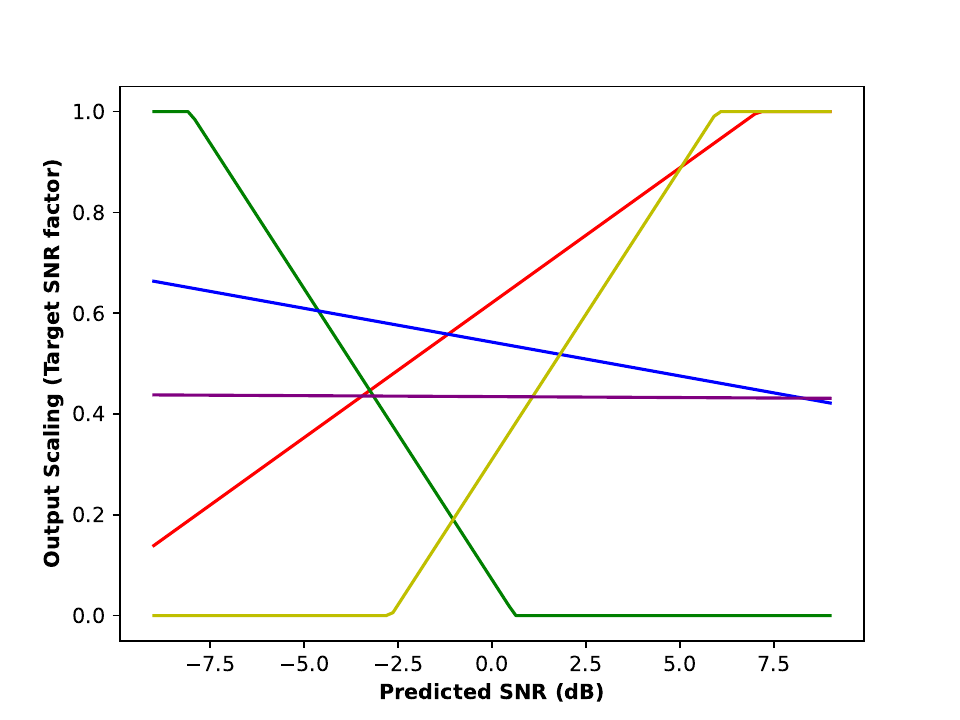}
    \includegraphics[scale=0.5]{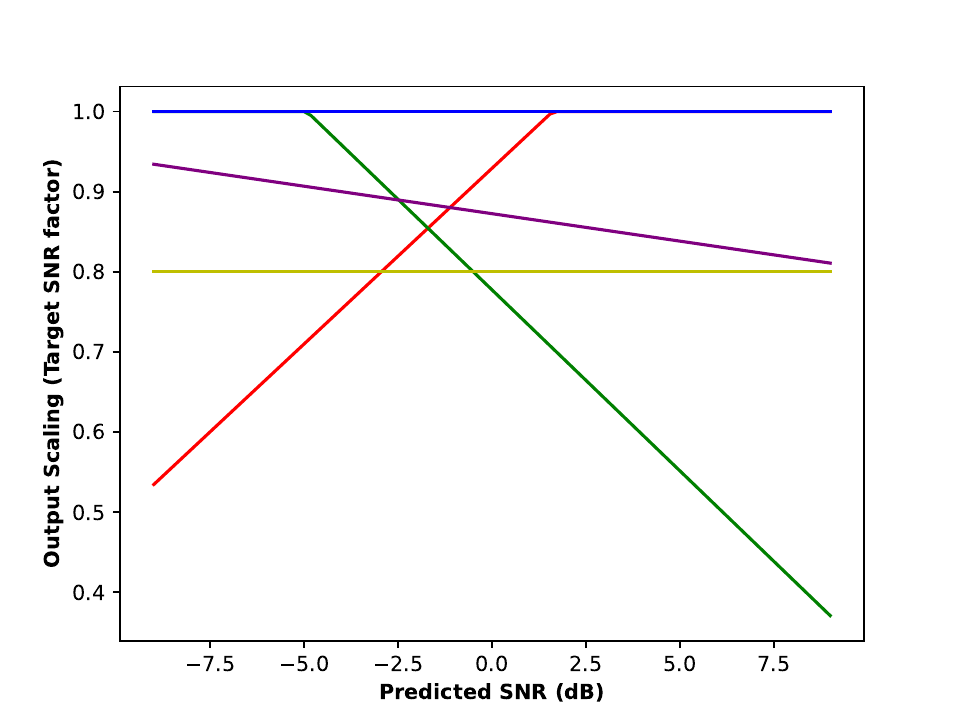}
    \caption{From left to right, top to bottom, HI-1 to HI-3's environmental preference functions. Key: Pedestrian-Red; Street-Green, Cafe-Blue, Bus-Yellow; Mean Function-Purple. The first participant shows preference for higher NR as the SNR increases, which indicates a dislike for the SE in general. For HI-2, the mixed gradients show a more situation specific preference, in the case of the yellow 'bus' environment the participant shows preference for no speech enhancement on the low SNR's: this could indicate a disdain for unneccesary artifacts as the noise itself is largely free from babble and has a stationary noise of the bus rumble. Interestingly this is not the case however for the green 'street' environment where the opposite is true, (similarly for the cafe), this could be because of the lower SNR's being less intelligible. For HI-3, as the hearing loss is severe, there is preference for almost maximum NR in most cases; the exceptions being low SNR Pedestrian (as this is where the experiment begins at 50\%, it is likely to do with the initial adjustment period of the experiment). The other exception is likely due to the street scene being more intelligible at high SNR's (a large open space with little babble or speech shaped noise), and the participant preferring a more natural sound -- this explanation is supported by HI-2 and the NH participants as well.} 
    \label{fig:multi-env-pref-plots-hi}
\end{figure}

Figure \ref{fig:hi_se_res} shows the HI results from the evaluation stage of the experiments.

Preliminary results show an improvement in speech quality for the preference learning vs. the enhanced condition in mild / moderate hearing loss participants HI-1 \& HI-2, the intelligibility also increases for HI-3 for the MaxSE condition (2) which shows that the enhancement has some benefit, however this is lost slightly with the introduction of PLSE: it is thought this may be because of the initial adjustment phase (starting at 50\% for 'Pedestrian' scene) not leading to preference congruent with the user's needs.

%\begin{table}%[hspace]
%\begin{tabular}{llll}
%Participant   & Speech Quality (mean +- std)        & Noise Quality                        & Intelligibility    \\
%HI-1(C1, C2) \Delta & (2.25 +- 1.30, 2.40 +- 1.36), +0.15   & (2.1 +- 1.30, 2.40 +- 1.36) +0.3    & (45\%+-30\%, 50\%+-29\%) +5\%  \\
%HI-2(C1, C2) \Delta & (2.5 +- 1.36, 3.45 +- 1.53),  +0.95 * & (3.1 +- 1.14, 4.25 +- 0.99), +1.15 ** & (60\%+-37\%, 72\%+-28\%), +12\% \\
%HI-3(C1, C2) \Delta & (2.9 +- 1.41, 2.55 +- 1.28), -0.35    & (2.9 +- 1.09, 2.4 +- 1.11), -0.5     & (48\%+-44\%, 45\%+-37\%) -3\%  \\
%HI-4(C1, C2) \Delta & (2.85 +- 1.33, 2.95 +- 1.41), +0.1     &  (1.98 +- 0.72, 2.93 +- 1.25), +0.95 **                                    & (58\%+-39\%, 49\% +-39\%) -9\%                    
%\end{tabular}
%\caption{\label{hi_results}Hearing Impaired Results, * indicates significance $p<0.05$ and ** indicates $p < 0.01$}. Note that for HI-4, the slight (non-significant) decrease in intelligibility score may be explained by a training effect as C2 preceded C1 and the participant was a non-native speaker, a similar effect was also observed in \cite{jkw-amhat2023}
%\end{table}

\section{Discussion, Conclusion and Future Work}

\subsection{Discussion and Conclusion}

In this paper, we have introduced a novel context-aware preference eliciation system for DNN based SE (particularly multi-modal, AV-SE). We presented it's multi-task, mask modulating formulation, and provided initial experimentation with NH and HI subjects. Though preliminary results show improvement for speech quality and perhaps intelligibility for the PLSE over the noisy and maxSE conditions, there are some limiting factors, most notably the small N of participants. Addtionally, the single-task nature of the experiments --- such as listening to the person speaking, though not being motivated by any background sounds -- limits the elicitation for preference of naturalness vs synthetic sound; however, this can be seen as a positive limitation in terms of disambiguating the analysis somewhat.

Next we visit each of these limitations in detail with respect to current and future work.

\subsection{Limitations and Future Work}
\subsubsection{Ambient Occlusion}
          \cite{ear-occluded-intel} seems to find that intelligibility improves greatly when there is more sound occluded via the ear canal. As such, it may be necessary to include ambient sound in the testing / evaluation chain whereby the original signal is summed with the enhanced signal and play this back in a quiet environment / with noise cancelling headphones for testing / evaluation according to a coefficient which simulates the amount of simulated Real-Ear-Insertion-Gain (to allow for different fittings).

            %\item We have purchased and began initial sensor readings for CL prediction with the the HP Reverb G2 Omnicept Edition VR headset.
            
            %The analysis of SE with respect to CL and possible integration in DNN SE model training / fine-tuning.
            %This will act as a natural extension to my preference learning / RL work and UoW's L5PC model in \textbf{Project Y3/Y4}.

\subsubsection{Extending To Unseen Acoustic Scenes}
In the case of environmental SNR, whilst the linear functions could be assumed to generalise to extreme SNRs that are not evaluated in this set up (though this should be tested), there is no reason to assume that the set of acoustic scenes here completes the set that would be necessary to provide the user with sufficient contextual output adaptation of the speech enhancement system. To extend and generalise the model, Bayesian modelling or fuzzy logic could be used to extend the predictive model to unseen acoustic scenes. A feature representation such as those shown in fig \ref{fig:tsne1} could be used to provide a low dimensional input that would effectively allow the scenes to be generalised like so (... is this neccessary or appropriate?)

\subsubsection{Extending to more than one target speaker}

A limitation of the study, and perhaps SE systems in general, is that they do not generalise to situations where there is more than one target speaker. In real life situations there is often more than one person we are conversing with. This introduces many complications, but it is of interest in future work to test if preference can be elicited in order to maximise intelligibility from more than one target speaker in a way that is suitable for hearing aids / hearing aid users. In a simple case, it may be important that a listener is able to hear their name being called whilst they are listening to a monologue.

\subsubsection{Refining The Preference Elicitation Framework}
The elicitation framework may not be perfect, and as with many optimisation problems the goal can be framed as balancing exploitation with exploration (cite?). The initial setting exploration phase may be particularly problematic (i.e. setting to 50 \% and expecting the user to adjust this sufficiently in a short number of utterances). Additionally, where the data is being collected for regression even though the user has entered a new environment with a particular setting that was previously suitable somewhere else. This could be improved by utilising an alternative AB test paradigm similar to \cite{nielsen2013hearing}, that utilises similar presumptions about function linearity but takes into account uncertainty into the preferences, given low data or high variance in particular acoustic conditions.

\subsubsection{Clinical Analysis}
Additionally, users will be able to visualise how their preference differs after the session according to a low-dimensional representation of the acoustic scenes. This will give the user and the audiologist some impression about how much noise suppression the user prefers and how that differs in various environments. It should also help to give the audiologist information on whether the user is a `noise hater' or a `distortion hater' \cite{pref-strength-individual-noisevsdistort} and use this to guide fittings, make HA recommendations and steer fine-tuning.

\subsubsection*{Acknowledgements}

This work is supported by the UK Engineering and Physical Sciences Research Council (EPSRC) programme grant: COG-MHEAR (Grant reference EP/T021063/1). 

We would like to thank Prof. Michael Akeroyd of University of Nottingham, School of Medicine for his insightful comments and feedback on the write up.

%\begin{figure}
%    \centering
%    \includegraphics[scale=0.4]{}
%    \caption{Screenshot from ASPIRE Dataset}
%    \label{fig:enter-label}
%\end{figure}

\bibliographystyle{apalike}
\bibliography{refs}

\begin{thebibliography}{}

\bibitem[Abdullah et~al., 2022]{IntegratingFA}
Abdullah, B.~M., Mobius, B., and Klakow, D. (2022).
\newblock Integrating form and meaning: A multi-task learning model for acoustic word embeddings.
\newblock In {\em Interspeech}.

\bibitem[Abeßer, 2020]{dl_asc_review}
Abeßer, J. (2020).
\newblock A review of deep learning based methods for acoustic scene classification.
\newblock {\em Applied Sciences}, 10(6).

\bibitem[Afouras et~al., 2018]{afouras2018conversation}
Afouras, T., Chung, J.~S., and Zisserman, A. (2018).
\newblock The conversation: Deep audio-visual speech enhancement.
\newblock {\em arXiv preprint arXiv:1804.04121}.

\bibitem[Andersen et~al., 2021]{creating-clarity-dl}
Andersen, A., Santurette, S., Pedersen, M., Alickovic, E., Fiedler, L., Jensen, J., and Behrens, T. (2021).
\newblock Creating clarity in noisy environments by using deep learning in hearing aids.
\newblock {\em Seminars in Hearing}, 42:260--281.

\bibitem[Barker et~al., 2015]{chime-3}
Barker, J., Marxer, R., Vincent, E., and Watanabe, S. (2015).
\newblock The third ‘chime’speech separation and recognition challenge: Dataset, task and baselines.
\newblock In {\em 2015 IEEE Workshop on Automatic Speech Recognition and Understanding (ASRU)}, pages 504--511. IEEE.

\bibitem[Bhat et~al., 2018]{Bhat2018}
Bhat, G.~S., Reddy, C.~K., Shankar, N., and Panahi, I.~M. (2018).
\newblock Smartphone based real-time super gaussian single microphone speech enhancement to improve intelligibility for hearing aid users using formant information.
\newblock In {\em 2018 40th Annual International Conference of the IEEE Engineering in Medicine and Biology Society (EMBC)}, pages 5503--5506. IEEE.

\bibitem[Brons et~al., 2012]{percep-fx-nr}
Brons, I., Houben, R., and Dreschler, W. (2012).
\newblock Perceptual effects of noise reduction with respect to personal preference, speech intelligibility, and listening effort.
\newblock {\em Ear and hearing}, 34.

\bibitem[Casolani et~al., 2024]{casolani2024evaluation}
Casolani, C., Borhan-Azad, A., S{\o}rensen, R.~S., Schlittenlacher, J., and Epp, B. (2024).
\newblock Evaluation of a fast method to measure high-frequency audiometry based on bayesian learning.
\newblock {\em Trends in Hearing}, 28:23312165231225545.

\bibitem[Chen et~al., 2018]{Chen2018}
Chen, J., Moore, B., Baer, T., and Wu, X. (2018).
\newblock Individually tailored spectral-change enhancement for the hearing impaired.
\newblock {\em The Journal of the Acoustical Society of America}, 143:1128--1137.

\bibitem[Chern et~al., 2023]{chern2023audio}
Chern, I.-C., Hung, K.-H., Chen, Y.-T., Hussain, T., Gogate, M., Hussain, A., Tsao, Y., and Hou, J.-C. (2023).
\newblock Audio-visual speech enhancement and separation by utilizing multi-modal self-supervised embeddings.
\newblock In {\em 2023 IEEE International Conference on Acoustics, Speech, and Signal Processing Workshops (ICASSPW)}, pages 1--5. IEEE.

\bibitem[Cooke et~al., 2006]{cooke2006audio}
Cooke, M., Barker, J., Cunningham, S., and Shao, X. (2006).
\newblock An audio-visual corpus for speech perception and automatic speech recognition.
\newblock {\em The Journal of the Acoustical Society of America}, 120(5):2421--2424.

\bibitem[Drakopoulos and Verhulst, 2022]{drakopoulos_differentiable_2022}
Drakopoulos, F. and Verhulst, S. (2022).
\newblock A {Differentiable} {Optimisation} {Framework} for {The} {Design} of {Individualised} {DNN}-based {Hearing}-{Aid} {Strategies}.
\newblock In {\em {ICASSP} 2022 - 2022 {IEEE} {International} {Conference} on {Acoustics}, {Speech} and {Signal} {Processing} ({ICASSP})}, pages 351--355, Singapore.

\bibitem[Esposito et~al., 2023]{semantically-informed-deep-neural-networks}
Esposito, M., Valente, G., Plasencia-Calaña, Y., Dumontier, M., Giordano, B.~L., and Formisano, E. (2023).
\newblock Semantically-informed deep neural networks for sound recognition.
\newblock In {\em ICASSP 2023 - 2023 IEEE International Conference on Acoustics, Speech and Signal Processing (ICASSP)}, pages 1--5.

\bibitem[Feng and Chen, 2022]{feng_nonintrusive_2022}
Feng, Y. and Chen, F. (2022).
\newblock Nonintrusive objective measurement of speech intelligibility: {A} review of methodology.
\newblock {\em Biomedical Signal Processing and Control}, 71:103204.

\bibitem[Fereczkowski et~al., 2024]{fereczkowski2024amplitude}
Fereczkowski, M., Sanchez-Lopez, R.~H., Christiansen, S., and Neher, T. (2024).
\newblock Amplitude compression for preventing rollover at above-conversational speech levels.
\newblock {\em Trends in Hearing}, 28:23312165231224597.

\bibitem[Hussain et~al., 2021]{io-avse}
Hussain, T., Gogate, M., Dashtipour, K., and Hussain, A. (2021).
\newblock Towards intelligibility-oriented audio-visual speech enhancement.
\newblock In {\em The Clarity Workshop on Machine Learning Challenges for Hearing Aids (Clarity-2021)}.

\bibitem[Jürgens et~al., 2023]{ear-occluded-intel}
Jürgens, T., Ihly, P., Tchorz, J., Zaar, J., Laugesen, S., Jones, G., and Santurette, S. (2023).
\newblock Individual real-ear occluded measurements as a predictor of speech intelligibility benefit from noise reduction in hearing aids.

\bibitem[Kates and Arehart, 2022]{haspi-hasqi}
Kates, J.~M. and Arehart, K.~H. (2022).
\newblock An overview of the haspi and hasqi metrics for predicting speech intelligibility and speech quality for normal hearing, hearing loss, and hearing aids.
\newblock {\em Hearing Research}, 426:108608.

\bibitem[Keidser et~al., 2011]{keidser2011nal}
Keidser, G., Dillon, H., Flax, M., Ching, T., and Brewer, S. (2011).
\newblock The nal-nl2 prescription procedure.
\newblock {\em Audiology research}, 1(1):e24.

\bibitem[Kirton-Wingate et~al., 2023]{jkw-amhat2023}
Kirton-Wingate, J., Ahmed, S., Gogate, M., Tsao, Y., and Hussain, A. (2023).
\newblock Towards individualised speech enhancement: An snr preference learning system for multi-modal hearing aids.
\newblock pages 1--5.

\bibitem[Kubiak et~al., 2022]{kubiak_relation_2022}
Kubiak, A.~M., Rennies, J., Ewert, S.~D., and Kollmeier, B. (2022).
\newblock Relation between hearing abilities and preferred playback settings for speech perception in complex listening conditions.
\newblock {\em International Journal of Audiology}, 61(11):965--974.

\bibitem[Lesica, 2018]{LESICA2018174}
Lesica, N.~A. (2018).
\newblock Why do hearing aids fail to restore normal auditory perception?
\newblock {\em Trends in Neurosciences}, 41(4):174--185.

\bibitem[Lo et~al., 2019]{mosnet}
Lo, C.-C., Fu, S.-W., Huang, W.-C., Wang, X., Yamagishi, J., Tsao, Y., and Wang, H.-M. (2019).
\newblock {MOSNet: Deep Learning-Based Objective Assessment for Voice Conversion}.
\newblock In {\em Proc. Interspeech 2019}, pages 1541--1545.

\bibitem[Mart{\'\i}n-Morat{\'o} et~al., 2022]{martin2022low}
Mart{\'\i}n-Morat{\'o}, I., Paissan, F., Ancilotto, A., Heittola, T., Mesaros, A., Farella, E., Brutti, A., and Virtanen, T. (2022).
\newblock Low-complexity acoustic scene classification in dcase 2022 challenge.
\newblock {\em arXiv preprint arXiv:2206.03835}.

\bibitem[Mermelstein, 1979]{segsnr}
Mermelstein, P. (1979).
\newblock {Evaluation of a segmental SNR measure as an indicator of the quality of ADPCM coded speech}.
\newblock {\em The Journal of the Acoustical Society of America}, 66(6):1664--1667.

\bibitem[Michelsanti et~al., 2021]{michelsanti_overview_2021}
Michelsanti, D., Tan, Z.-H., Zhang, S.-X., Xu, Y., Yu, M., Yu, D., and Jensen, J. (2021).
\newblock An overview of deep-learning-based audio-visual speech enhancement and separation.
\newblock {\em IEEE/ACM Trans. Audio, Speech and Lang. Proc.}, 29:1368–1396.

\bibitem[Mitchell et~al., 2023]{mitchell2023deep}
Mitchell, A., Brown, E., Deo, R., Hou, Y., Kirton-Wingate, J., Liang, J., Sheinkman, A., Soelistyo, C., Sood, H., Wongprommoon, A., et~al. (2023).
\newblock Deep learning techniques for noise annoyance detection: Results from an intensive workshop at the alan turing institute.
\newblock {\em The Journal of the Acoustical Society of America}, 153(3\_supplement):A262--A262.

\bibitem[Moore et~al., 2022]{diag-noise-induced-hearing-loss}
Moore, B. C.~J., Lowe, D.~A., and Cox, G. (2022).
\newblock Guidelines for diagnosing and quantifying noise-induced hearing loss.
\newblock {\em Trends in Hearing}, 26.
\newblock PMID: 35469496.

\bibitem[Neher and Wagener, 2016]{Neher2016}
Neher, T. and Wagener, K. (2016).
\newblock Investigating differences in preferred noise reduction strength among hearing aid users.
\newblock {\em Trends in Hearing}, 20.

\bibitem[Nielsen et~al., 2014]{gp-personal}
Nielsen, J., Nielsen, J., and Larsen, J. (2014).
\newblock Perception-based personalization of hearing aids using gaussian processes and active learning.
\newblock {\em IEEE/ACM Transactions on Audio, Speech, and Language Processing}, 23.

\bibitem[Nielsen et~al., 2013]{nielsen2013hearing}
Nielsen, J.~B., Nielsen, J., Jensen, B.~S., and Larsen, J. (2013).
\newblock Hearing aid personalization.
\newblock In {\em 27th Annual Conference on Neural Information Processing Systems (NIPS 2013)}.

\bibitem[Panariello, 2022]{panariello2022low}
Panariello, M. (2022).
\newblock {\em Low-complexity neural networks for robust acoustic scene classification in wearable audio devices}.
\newblock PhD thesis, Politecnico di Torino.

\bibitem[Reinten et~al., 2023]{pref-strength-individual-noisevsdistort}
Reinten, I., de~Ronde-Brons, I., Houben, R., and Dreschler, W. (2023).
\newblock Individual listener preference for strength of single-microphone noise-reduction; trade-off between noise tolerance and signal distortion tolerance.
\newblock {\em Trends in Hearing}, 27:23312165231192304.

\bibitem[Tyagi and Rajan, 2022]{tsne_info}
Tyagi, A. and Rajan, P. (2022).
\newblock Location-invariant representations for acoustic scene classification.
\newblock In {\em 2022 30th European Signal Processing Conference (EUSIPCO)}, pages 394--398.

\bibitem[Wang et~al., 2020]{wang2020robust}
Wang, W., Xing, C., Wang, D., Chen, X., and Sun, F. (2020).
\newblock A robust audio-visual speech enhancement model.
\newblock In {\em ICASSP 2020-2020 IEEE international conference on acoustics, speech and signal processing (ICASSP)}, pages 7529--7533. IEEE.

\bibitem[Weber et~al., 2021]{weber-etal}
Weber, L., Jumelet, J., Bruni, E., and Hupkes, D. (2021).
\newblock Language modelling as a multi-task problem.
\newblock In {\em Proceedings of the 16th Conference of the European Chapter of the Association for Computational Linguistics: Main Volume}, pages 2049--2060, Online. Association for Computational Linguistics.

\bibitem[wei Fu et~al., 2018]{Quality-net}
wei Fu, S., Tsao, Y., Hwang, H.-T., and Wang, H.-M. (2018).
\newblock Quality-net: An end-to-end non-intrusive speech quality assessment model based on blstm.
\newblock In {\em Proc. Interspeech 2018}, pages 1873--1877.

\bibitem[Zezario et~al., 2022a]{mbi-net}
Zezario, R., Chen, F., Fuh, C.-S., Wang, H.-m., and Tsao, Y. (2022a).
\newblock Mbi-net: A non-intrusive multi-branched speech intelligibility prediction model for hearing aids.
\newblock In {\em Interspeech}.

\bibitem[Zezario et~al., 2022b]{multi-obj-speech-assess}
Zezario, R.~E., Fu, S.-W., Chen, F., Fuh, C.-S., Wang, H.-M., and Tsao, Y. (2022b).
\newblock Deep learning-based non-intrusive multi-objective speech assessment model with cross-domain features.
\newblock {\em IEEE/ACM Transactions on Audio, Speech, and Language Processing}, 31:54--70.

\bibitem[Zezario et~al., 2023]{mosa-net}
Zezario, R.~E., Fu, S.-W., Chen, F., Fuh, C.-S., Wang, H.-M., and Tsao, Y. (2023).
\newblock Deep learning-based non-intrusive multi-objective speech assessment model with cross-domain features.
\newblock {\em IEEE/ACM Transactions on Audio, Speech, and Language Processing}, 31:54--70.

\bibitem[Zezario et~al., 2020]{stoinet}
Zezario, R.~E., Fu, S.-W., Fuh, C.-S., Tsao, Y., and Wang, H.-M. (2020).
\newblock Stoi-net: A deep learning based non-intrusive speech intelligibility assessment model.
\newblock In {\em 2020 Asia-Pacific Signal and Information Processing Association Annual Summit and Conference (APSIPA ASC)}, pages 482--486. IEEE.

\bibitem[Zhang and Yang, 2022]{MTL}
Zhang, Y. and Yang, Q. (2022).
\newblock A survey on multi-task learning.
\newblock {\em IEEE Transactions on Knowledge and Data Engineering}, 34(12):5586--5609.

\bibitem[Zhao et~al., 2019]{Phone-Aware-MTL}
Zhao, M., Li, R., Yan, S., Li, Z., Lu, H., Xia, S., Hong, Q., and Li, L. (2019).
\newblock Phone-aware multi-task learning and length expanding for short-duration language recognition.
\newblock In {\em 2019 Asia-Pacific Signal and Information Processing Association Annual Summit and Conference (APSIPA ASC)}, pages 433--437.

\bibitem[Zheng et~al., 2023]{sixty-years-se}
Zheng, C., Zhang, H., Liu, W., Luo, X., Li, A., Li, X., and Moore, B. C.~J. (2023).
\newblock Sixty years of frequency-domain monaural speech enhancement: From traditional to deep learning methods.
\newblock {\em Trends in Hearing}, 27:23312165231209913.
\newblock PMID: 37956661.

\end{thebibliography}

\end{document}